\pdfoutput=1

\documentclass[aps,pra,twocolumn,superscriptaddress,floatfix]{revtex4-1}
\usepackage{amssymb}
\usepackage{amsmath}
\usepackage{graphicx}
\usepackage{xcolor}
\usepackage{siunitx}

\begin{document}

\title{On the optimal design of grid-based binary holograms for matter wave lithography}

\author{Torstein Nesse}
\affiliation{Department of Physics, NTNU Norwegian University of Science and Technology, NO-7491 Trondheim, Norway}

\author{Jean-Philippe Banon}
\affiliation{Department of Physics, NTNU Norwegian University of Science and Technology, NO-7491 Trondheim, Norway}

\author{Bodil Holst}
\affiliation{Department of Physics and Technology, University of Bergen, All\'egaten 55, NO-5007 Bergen, Norway}

\author{Ingve Simonsen}
\email{ingve.simonsen@ntnu.no}
\affiliation{Department of Physics, NTNU Norwegian University of Science and Technology, NO-7491 Trondheim, Norway}
\affiliation{Surface du Verre et Interfaces, UMR 125 CNRS/Saint-Gobain, F-93303 Aubervilliers, France}

\date{\today}

\begin{abstract}
Grid based binary holography~(GBH) is an attractive method for patterning with light or matter waves. It is an approximate technique in which different holographic masks can be used to produce similar patterns. Here we present an optimal design method for GBH masks that allows for freely selecting the fraction of open holes in the mask from below 10\% to above 90\%. Open-fraction is an important design parameter when making masks for use in lithography systems. The method also includes a rescaling feature that potentially enables a better contrast of the generated patterns. Through simulations we investigate the contrast and robustness of the patterns formed by masks generated by the proposed optimal design method. It is demonstrated that high contrast patterns are achievable for a wide range of open-fractions. We conclude that reaching a desired open-fraction is a trade-off with the contrast of the pattern generated by the mask.
\end{abstract}


\maketitle


\section{Introduction}
High resolution lithography is a central part of the semiconductor industry. Industrial lithography is usually mask based, to ensure a high production speed. In mask based lithography a photonic beam projects a direct image of a mask onto a substrate. By superposition of images from several masks combined with etching and metal deposition processing steps, micro-chips are created. The lithography resolution is one of the most important parameters in determining how small integrated circuits can be made (thickness of wires etc.) and the size of the circuits ultimately determines the speed of the electronic components~\cite{Borkar11}.

In standard photolithography the wavelength of the light being used determines the resolution: the smaller the wavelength, the higher the resolution. The present industrial photolithography standard is the immersion scanner using a \SI{193}{nm} light source in a high numerical aperture medium~\cite{ITRS}. The industry is currently implementing the next generation of lithography devices, extreme ultraviolet lithography (EUV) based on a \SI{13.5}{nm} wavelength light source, which together with immersion techniques and multiple patterning is expected to be able to scale down to critical dimensions below \SI{14}{\nano\meter}~\cite{ITRS}.

Atom lithography has been suggested as a possible future step for high resolution lithography. The de Broglie wavelength of thermal atoms is much smaller than the wavelength of optical photons. For helium atoms at thermal energies between \SI{20}{meV} and \SI{60}{meV}, the corresponding wavelength is \SI{0.1}{nm} or less. This makes atom beams in principle a very attractive candidate for pattern generation. One approach in atom lithography is to use a beam of metastable atoms for the pattern generation~\cite{Berggren95,Baldwin05}. When an atom hits the substrate it decays and the energy of the metastable state is transferred to the substrate. Various difficulties have prevented atom lithography from being used industrially. A major problem has been to create coherent atom beams. This will be discussed in more details later in this section. Another problem is that low energy metastable atoms do not penetrate solid materials, so it is not clear how mask based atom lithography can be performed.



This problem was in principle already solved more than 20 years ago, using binary holography. Binary holography was originally developed by Lohmann and Paris~\cite{Lohmann67} to shape incident electromagnetic beams and it uses masks made out of a set of regions that are either completely transparent or opaque. Originally the method was developed to create holograms for electromagnetic waves using a computer, and the procedure is often referred to as computer generated holography~(CGH) in the literature. Due to the de~Broglie-wavelength associated with a matter wave, the method also works for atom beams. Using the method, an approximation of any arbitrary pattern can be generated on a substrate by using a mask that represents a Fourier transform of the desired pattern. The work of Lohmann and Paris was used by Onoe and Kaneko~\cite{Onoe79} to develop grid based binary holography (GBH). In GBH the masks are based on a grid of identical holes that are either completely open or closed. Masks produced this way can also be used to approximate arbitrarily shaped beams.

Using GBH Fuijta~\textit{et al.}~\cite{Fujita96} successfully created patterns using a beam of metastable Ne atoms and a silicon nitride mask with \SI{30}{nm} holes created by electron lithography and reactive ion etching. Provided the incident beam is coherent and of a wavelength significantly smaller than the grid period, the resolution limit in GBH is given by the grid period. With present day electron lithography technology, patterns can be made at a resolution of \SI{5}{nm} or less~\cite{Manfrinato13}, which means that atom based lithography in the sub-\SI{10}{nm} regime could be possible. However, this still has to be demonstrated.

After the work of Fujita~\textit{et al.}~\cite{Fujita96}, several experiments have been performed on the manipulation of atom and molecular beams with ``optical'' elements created out of free standing, solid structures. In 2012, a Fresnel zone plate etched into a silicon nitride membrane was used to focus a neutral helium beam down to a spot size of sub-micron diameter~\cite{Eder12}. In this study it was found that the resolution was limited by the velocity spread of the beam~\cite{Eder12}. Further work by the same group has shown focusing with a photon-sieve structure~\cite{Eder15}. More recently a diffraction grating for molecular interferometry was created out of a graphene membrane~\cite{Arndt15}~(see also~\cite{Pritchard09}). This year helium diffraction from a micron scale structure was achieved for the first time~\cite{Nesse17}.


As mentioned above, the main problem for atom lithography has been to create coherent beams. Significant progress has been made here within the last few years. The spatial coherence of a molecular beam is determined by the velocity distribution~\cite{Patton2006}. Recently a beam of metastable helium atoms with an average wavelength of around \SI{0.06}{\nano\meter} and a very narrow wavelength distribution $\lambda/\delta\lambda = 200$, was produced using a pulsed source~\cite{Even2015}.

The ultimate coherent beam would seem to be a Bose Einstein condensate (BEC). A BEC of metastable atoms, which brings an assembly of atoms into exactly the same energy state, was created some years back~\cite{Robert01,Santos01}. Recently, Zeilinger~\textit{et al.} generated a beam of BEC metastable helium atoms; that is, a perfectly coherent beam of metastable atoms~\cite{Zeilinger14}. However, the Fraunhofer diffraction formula does not apply for a BEC~\cite{Fouda2016}. Furthermore, the de Broglie wavelength of a dropping BEC of helium is very large, about \SI{30}{\nano\meter} after a drop of half a meter. For high resolution lithography one wants to use small wavelengths. One can think of experimental ways to get around this, for example by moving the mask relative to the BEC so that the BEC wavelength relative to the mask gets smaller, but in any case, considerable amendments would have to be made to the theory we present here.


The holographic structure of the mask means that the influence of local mask errors is less prominent in the final pattern. This makes GBH interesting for photon and mask based electron lithography. Furthermore, the exposed areas are more evenly distributed across the whole mask which makes it thermally more stable. This is particularly an issue in EUV lithography. For photonic lithography applications the hole structure can be placed on a suitable substrate.

In this paper the work of Onoe and Kaneko~\cite{Onoe79} is further developed. We investigate the contrast and robustness of the patterns created by the hologram masks and provide an algorithm that can be used to select the open hole fraction of a mask over a large range. From a mask fabrication point of view the most desirable is a solution that minimizes the number of holes, but from a chip fabrication point of view it may be better to maximize the number of holes, in order to reduce the heating of the mask. Similar to Omoe and Kaneko we work in the Fraunhofer diffraction limit. For a real appliation a lens may need to be introduced to bring the pattern closer and to increase resolution. Several experiments have been carried out focussing molecular beams using zoneplates~\cite{Eder2017,Eder15,Koch2008,Doak1999,Carnal1991} or electromagnetic~\cite{Gardner2017} lenses and this principle could be applied here.

\smallskip
The rest of this paper is organized as follows; First a presentation of the theoretical framework of GBH followed by a discussion of the specific approach to mask generation is given, then the new method that allows for an arbitrary choice of the open-fraction of the mask is presented. We also look at parts of the mask generation method that hasn't previously been discussed and highlight parameters that are important to the properties of the final mask. The method is then applied to some illustrative examples. Finally we explore the contrast of masks made with different open hole fractions, and characterize the robustness of the masks.

\section{\label{sec:GBBH}Grid-based binary holograms}

%
\begin{figure*}
\centering\includegraphics[width=1.2\columnwidth]{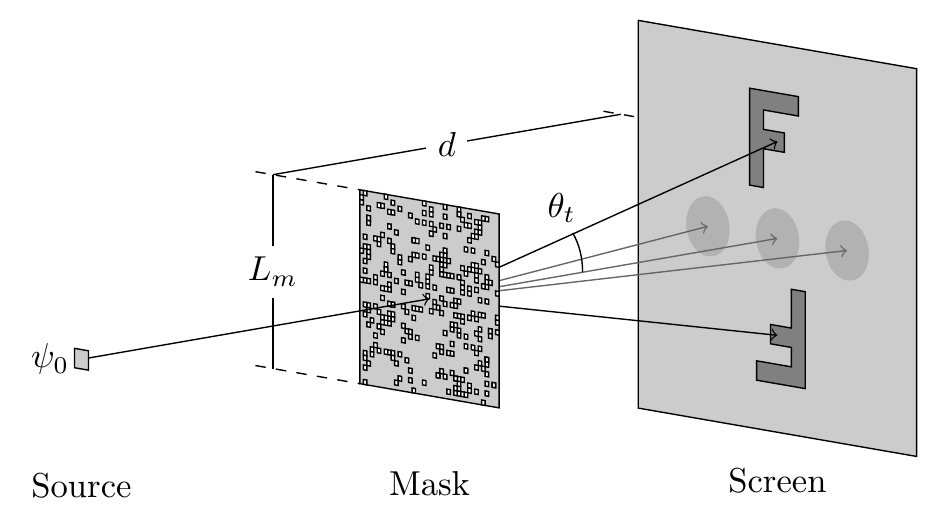}
\caption{\label{fig:system}Diagram of the system setup. Plane waves are entering the binary mask at normal incidence. The desired pattern is produced at an angle of transmission $\theta_t$ away from the normal direction in the screen plane. This is one way to encode different phases. The central circle represent the zeroth order diffraction peak, and the two other circles represent the first order diffraction peaks in the perpendicular direction to our patterns. These locations are not encoded for in the mask and show random intensity distributions. In a lithography setup they could be blocked before hitting the screen.
}
\end{figure*}

\subsection{Theoretical formulation}
Holograms are constructed to impose a given amplitude and phase on different parts of a field interacting with them~\cite{Goodman05}. A binary hologram is a special type of hologram where the mask is binary, that is, it is made from parts that are either completely open or closed to the incident field. When the open and closed sections of the binary mask form a grid, then one talks about GBH.

Consider the geometry depicted in Fig.~\ref{fig:system}; here a beam is incident normally onto a binary mask and a pattern is observed on a screen located a distance $d$ behind it. In the following, it will be assumed that the front and back surfaces of the mask and the surface of the screen all are planar and parallel. A coordinate system is defined so that the back surface of the mask coincides with the plane $x_3 = 0$ and the screen is located a distance $x_3 = d > 0$ behind it. The incident beam is modeled by the incident plane scalar wave
\begin{equation}
	\psi_0(\mathbf{x}) = \exp(\mathrm{i}\mathbf{k}\cdot\mathbf{x}) = \exp(\mathrm{i}kx_3),
\end{equation}
where the incident wave vector is given by $\mathbf{k} = k\hat{\mathbf{x}}_3$, with $k = 2\pi/\lambda$ where $\lambda$ denotes the wavelength of the incident beam; in the case of a matter wave (atoms, molecules or electrons), $\lambda$ represents the de Broglie wavelength. The presence of the mask will transform the incident field at the front side of the mask into the ``mask field'' 
\begin{align}
  \psi_M(\mathbf{x}_\|) 
  &= 
  T \psi_0(\mathbf{x})\rvert_{x_3 = -\tau},
  \label{eq:psim}
\end{align}
at the back side of the mask where $\tau\ll d$ is the thickness of the mask and $\mathbf{x}_\| = (x_1,x_2,0)$ represents a point in the $x_1x_2$ plane --- the ``mask plane''. In writing Eq.~\eqref{eq:psim}, we have introduced the transfer operator (or function), $T$, that encodes the details of the mask, such as its finite thickness and binary nature, and relates the field at the front side of the mask to the field at the back side of the mask.

In the simplest case, we have an ideal mask with a small thickness $\tau$ so that the operator $T$ equals the binary function
\begin{align}
  T_m(\mathbf{x}_\|) 
  &=
  \begin{cases}
    1 & \text{ if }\mathbf{x}_\| \in \text{hole} \\
    0 & \text{ otherwise}
  \end{cases}.
\end{align}
When it is applied to the incoming field at $x_3=-\tau\approx 0$ one gets
\begin{align}
  \psi_M(\mathbf{x})\rvert_{x_3\approx 0}
  = \psi_M(\mathbf{x}_\|\rvert 0) 
  = T_m(\mathbf{x}_\|)\psi_0(\mathbf{x}_\|).
  \label{eq:psim2}
\end{align}

\begin{widetext}

Let $\mathbf{x}'_\parallel$ denote a point on the surface of the screen $x_3=d$. Following standard diffraction theory, the field in the screen plane can be written as $[\psi_s(\mathbf{x}')\rvert_{x_3=d} = \psi_s(\mathbf{x}'_\|\rvert d)]$~\cite{Goodman05,Adams94}
\begin{align}
  \psi_s(\mathbf{x}'_\|\rvert d) 
  &=
  -\frac{\mathrm{i}}{2\pi}
  \frac{k}{d}
  \exp{ (\mathrm{i}kd) }
  \exp{ \left( \frac{\mathrm{i}}{2}\frac{k}{d} x_\parallel'^2 \right) }
  \int\mathrm{d}^2x_\parallel\, 
  \psi_M(\mathbf{x}_\parallel |  0)
  \exp{
    \left(
      -\mathrm{i}\frac{k}{d} \mathbf{x}'_\|  \cdot \mathbf{x}_\| 
    \right)},
  \label{eq:fraunhofer}
\end{align}
which is the Fraunhofer zone limit of the Rayleigh-Sommerfeld diffraction formula. This expression is valid when the screen plane is many wavelengths away from the mask plane ($d\gg\lambda$)~\cite{Goodman05}. A quantitative criterion for the validity of Eq.~\eqref{eq:fraunhofer} is $d > kL_m^2/\pi$ where  $L_m$ represents the lateral extent of the mask~\cite{Goodman05}. To obtain the field closer to the mask, the Fresnel approximation or the full Rayleigh-Sommerfeld diffraction formula can be employed. In the current work we will only use the Fraunhofer approximation because of its simplicity, but the GBH method presented later can also be valid for mask fields found by other methods.

Equation~\eqref{eq:fraunhofer} together with Eq.~\eqref{eq:psim2} makes it possible to find the response from a binary mask at a plane a long distance behind the mask. However, what we really want is to construct a desirable mask for a given final pattern in the screen plane (see Fig.~\ref{fig:system}). Therefore we now turn the problem around, and find an expression for the field just behind the mask given a target intensity in a screen plane far behind the mask. We call this field $\psi_m(\mathbf{x}_\parallel|0)$ to differentiate it from the mask field $\psi_M(\mathbf{x}_\parallel|0)$ created by the incident field and the transfer function of the binary mask given by Eq.~\eqref{eq:psim2}. In order to determine the unknown mask field $\psi_m(\mathbf{x}_\parallel|0)$ from the knowledge of the field in the screen plane, $\psi_s(\mathbf{x}_\parallel'| d)$, one starts by noting that the integral that appears in Eq.~\eqref{eq:fraunhofer} is the  Fourier transform of  $\psi_m(\mathbf{x}_\parallel|0)$ evaluated at the wave vector $\mathbf{K}_\parallel=(k/d)\mathbf{x}_\parallel'$. By taking the inverse Fourier transform of both sides of Eq.~\eqref{eq:fraunhofer} and using that $\mathrm{d}^2K_\parallel = (k/d)^2 \mathrm{d}^2x_\parallel'$, one readily finds

\begin{align}
  \psi_m(\mathbf{x}_\parallel | 0) 
  &=
  \frac{\mathrm{i}}{2\pi}
  \frac{k}{d}
  \exp{\left(-\mathrm{i}kd\right)}
  \int\mathrm{d}^2{x}_\parallel'\,
  \exp \left(-\frac{\mathrm{i}}{2} \frac{k}{d} x_\parallel'^2 \right)
  \psi_s(\mathbf{x}_\parallel'\rvert d)
  \exp{\left(
      \mathrm{i}\frac{k}{d} \mathbf{x}'_\parallel \cdot \mathbf{x}_\parallel 
    \right)}.
  \label{eq:inverse_fraunhofer}
\end{align}

\end{widetext}
Equation~\eqref{eq:inverse_fraunhofer} states that the mask field $\psi_m$ is the inverse Fourier transform of the field in the screen plane $\psi_s$ times a propagating factor depending on the mask-screen separation.

%
\begin{figure}
\centering\includegraphics[width=0.8\columnwidth]{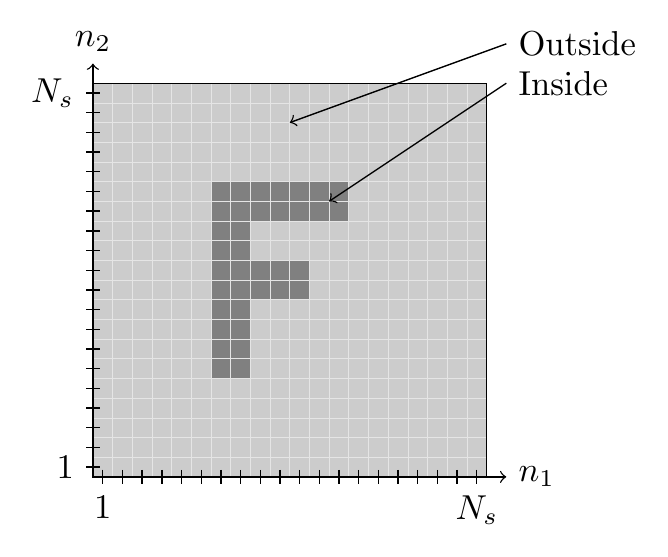}
\caption{\label{fig:target} Example of a discretized target pattern with $N_s~\times~N_s$ points. These points are indexed from $1$ to $N_s$ in each direction, which corresponds to real space coordinates given by Eq.~\eqref{fig:target_coordinates}.
}
\end{figure}

%
\begin{figure}
  \centering
  \includegraphics[width=\columnwidth]{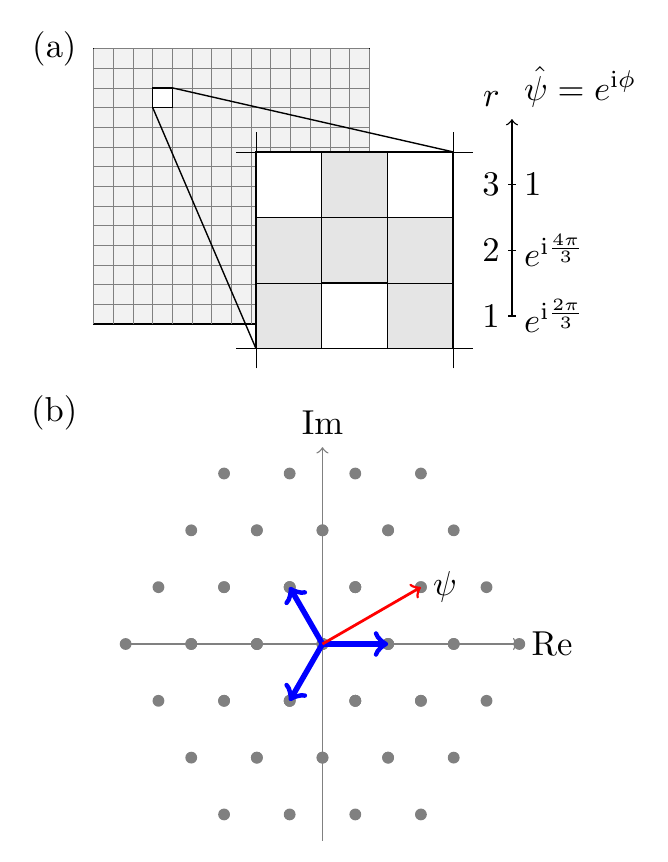}
  \caption{(Color online)(a) Illustration of the subcell-structure with three open subcells (white cells) assuming $S=3$. The different rows have different phases $\phi = (2\pi r)/S$, with $r=1,\mathellipsis,S$, associated with them. Each open subcell leads to a contribution to the overall field $\Psi_m$ from the cell in the direction $\exp(\mathrm{i}\phi)$. The field corresponding to the shown subcell therefore is $\Psi_m = \left[2+\exp\left({\mathrm{i}2\pi/3}\right)\right]/3$. This must be scaled with the incident field. (b) All points in the complex plane that can be realized by a configuration of a subcell. The thick vectors (blue) correspond to one open subcell from each row. The thin vector (red) represents the contribution to the field from the open subcells shown in Fig.~(a). The amplitude of the vectors must be scaled so that the available discretization region covers the amplitude range of the mask field $\Psi_m(\mathbf{n})$.}
  \label{fig:gridex}
\end{figure}

If the \textit{desired} intensity distribution in the screen plane $x_3=d$ is denoted $I_s(\mathbf{x}_\parallel'\rvert d)$, then the field in this plane can be written in the form [$I_s=|\psi_s|^2$] 
\begin{align}
  \label{eq:desired_screen_field}
  \psi_s(\mathbf{x}'_\parallel| d) 
  &= 
  \sqrt{I_s(\mathbf{x}'_\parallel | d)} \exp\left[\mathrm{i}\Phi(\mathbf{x}'_\| \rvert d)\right],
\end{align}
where $\Phi(\mathbf{x}'_\|\rvert d)$ is a random phase function that is independent of the choice made for the intensity $I_s$. Without loss of generality, we will in the following assume the phase function to be spatially uncorrelated and uniformly distributed on the interval $[0, 2\pi)$. An example target pattern is shown in Fig.~\ref{fig:target}. Two levels of gray are used to denote inside and outside of the pattern. In most cases the absolute intensity level of the target pattern $I_s(\mathbf{x}'_\parallel | d)$ is not important; the main concern is the contrast, that is the relative difference between the intensity inside and outside of the pattern. Therefore an intensity rescaled image serves as our target pattern for the GBH.

As a result of Eq.~\eqref{eq:desired_screen_field}, the field $\psi_m(\mathbf{x}_\parallel | 0)$ will have a complex phase variation over the (back) surface of the mask. Ideally, the field $\psi_M(\mathbf{x}_\parallel | 0 )$ that is related to the incident field via Eq.~\eqref{eq:psim2}, should equal $\psi_m(\mathbf{x}_\parallel | 0 )$. However, at normal incidence, this is not possible since the operator $T_m$ does not change the phase of the field on which it operates, only its amplitude. In order to allow for a changing phase in $\psi_M(\mathbf{x}_\parallel | 0 )$ over the surface of the mask, one could, for instance, assume that the field of incidence is impinging \textit{non-normally} ($\theta_0\neq\ang{0}$)  onto this surface. With a phase change along $\psi_0(\mathbf{x}_\parallel)$, a changing phase is available for $\psi_M(\mathbf{x}_\parallel | 0 )$.

An alternative (but equivalent) approach is to keep $\theta_0=\ang{0}$ for the incident field, but form the image in a region of the screen plane that corresponds to a \textit{non-zero} polar angle of transmission $\theta_t$. This situation is presented in Fig.~\ref{fig:system}. In the former approach the phase difference is due to the difference in propagation path of the incident field, while in the latter approach, it is due to the path difference of the transmitted field. Hence, the two approaches are in principle equivalent but it should be noted that in the latter case $\psi_M(\mathbf{x}_\parallel | 0 )$ is different from $\psi_m(\mathbf{x}_\parallel | 0 )$. The latter approach has the advantage that if an image is formed in the region around the angles of transmission $(\theta_t,\phi_t)$, where $\phi_t$ denotes the azimuthal angle of transmission, then a  similar image will be formed in the direction $(\theta_t,\phi_t+\ang{180})$ only rotated \ang{180} relative the first image; see Fig.~\ref{fig:system} (where $\phi_t=\pm\ang{90}$).

In this work we use the second approach since for a given mask structure it results in two (or more) images, and the case of normal incidence is usually more easy to handle experimentally. The details of this approach will be discussed in the following subsection.

It should be mentioned that a possible third alternative is a combination of the former two but this is technically more complicated and will not be discussed here.

\subsection{Numerical construction of GBHs}

Equations~\eqref{eq:inverse_fraunhofer} and \eqref{eq:desired_screen_field} allow us to calculate the mask field $\psi_m(\mathbf{x}_\parallel | 0)$ required to produce a given intensity in the screen plane. However, when using a binary mask, any general mask field cannot be realized with a desired (high) precision; instead only an approximate solution is expected and the purpose of this subsection is to describe how an approximate mask field can be constructed given the intensity distribution (the ``pattern'') in the screen plane [see Eq.~\eqref{eq:desired_screen_field}].

The procedure to construct the desired mask starts by discretizing the target intensity pattern $I_s(\mathbf{x}_\parallel' | d)$ in the screen plane. If this pattern in the screen plane  fits inside a square region of sides $L_s$, we discretize $I_s(\mathbf{x}_\parallel' | d)$ onto a grid of $N_s \times N_s$ points of coordinates
\begin{subequations}
\label{eq:screen_plane_discretization}
\begin{align}
  \mathbf{x}_\parallel'(\mathbf{n}) &= \big(x_1'(n_1), x_2'(n_2), 0 \big) 
\end{align} 
where ($i=1,2$)
\begin{align} 
	x_i'(n_i) &= -\frac{L_s}{2} + \left(n_i - \frac{1}{2}\right) \Delta x_s, 
	\label{fig:target_coordinates}
\end{align} 
\end{subequations}
with  $n_{i}=1,\ldots,N_s$ and $\Delta x_s = L_s/N_s$. See example target pattern in Fig.~\ref{fig:target}.

To obtain the discretization in the mask plane, one recalls from Eq.~\eqref{eq:inverse_fraunhofer} that  $\psi_m(\mathbf{x}_\parallel | 0)$  is expressed  as the inverse Fourier transform of a function containing $\psi_s(\mathbf{x}_\parallel'|d)$ [or $I_s(\mathbf{x}_\parallel'|d)$]. Hence, the discretization in the plane of the mask (direct space) is determined from the assumed discretization in the screen plane (Fourier space) [Eq.~\eqref{eq:screen_plane_discretization}]. Let $\Delta x_m$ and $L_m$ denote the discretization interval and spatial extent of direct space, respectively. Since the relevant Fourier variable is $\mathbf{K}_\parallel = (k/d)\mathbf{x}_\parallel'$, the Nyquist frequency and sampling interval (for both orthogonal directions) of Fourier space are given by~\cite{Book:NumericalRecipies} $K_\star=\pi/\Delta x_m= L_sk/(2d)$ and $\Delta K=2\pi/L_m = k\Delta x_s/d$, respectively. Thus, one is led to conclude that 
\begin{align}
  \label{eq:mask-plane-discretiatin-parameter}
  \Delta x_m &= \frac{d}{k} \frac{2\pi}{L_s},
  &
  L_m       &= N_s \Delta x_m,
\end{align}    
so the discretization of the mask plane becomes
\begin{subequations}
  \label{eq:mask_plane_discretization}
\begin{align}
  \mathbf{x}_\parallel(\mathbf{n}) &= \big( x_1(n_1), x_2(n_2), 0 \big), 
\end{align}
where
\begin{align}
  x_i(n_i) &= -\frac{L_m}{2} + \left(n_i - \frac{1}{2}\right) \Delta x_m.
\end{align}
\end{subequations}

With the use of Eqs.~\eqref{eq:screen_plane_discretization} and \eqref{eq:mask_plane_discretization}, the discrete version of Eq.~\eqref{eq:inverse_fraunhofer} can be written in the form
\begin{widetext}

\begin{align}
  \psi_m\left(\mathbf{x}_\parallel(\mathbf{n}) | 0 \right)
  &=
  \frac{\mathrm{i}}{2\pi}
  \frac{k}{d}
    \exp{\left(-\mathrm{i}kd\right)}
    \sum\limits_{\mathbf{n'}} \Delta{x}_s^2\,
  \exp \left(-\frac{\mathrm{i}}{2} \frac{k}{d}x_\parallel'^2(\mathbf{n}') \right)
  \psi_s(\mathbf{x}_\parallel'(\mathbf{n}') | d)
  \exp{\left(
      \mathrm{i}\frac{k}{d}  
      \mathbf{x}'_\|(\mathbf{n}')  \cdot  \mathbf{x}_\|(\mathbf{n})
    \right)}.
  \label{eq:discretized_mask}
\end{align}

\end{widetext}
In this and later equations it is implicitly understood that $\psi_s(\mathbf{x}_\parallel' | d)$ is given in terms of Eq.~\eqref{eq:desired_screen_field}.

Following the terminology previously introduced by Onoe and Kaneko~\cite{Onoe79}, a \textit{cell} is defined as the square region of the mask plane centered at $\mathbf{x}_\parallel(\mathbf{n})$ and having sides $\Delta x_m$. The aim is to create a scheme which, within the restrictions of binary masks, can be used to generate masks that approximately give rise to the desired mask field for each cell.

Several methods have been proposed in the literature for the purpose of approximating the field from a single cell. Here we adopt the method introduced by Onoe and Kaneko~\cite{Onoe79} for the construction of grid based binary holograms --- what they call \textit{pure binary holograms}.
This method operates on the principle of creating a \textit{subgrid} within each cell which structure encodes the desired $\psi_m$ that is associated with that cell for given well-defined angles of transmission $(\theta_t,\phi_t)$ to be defined below; cf.  Figs.~\ref{fig:system} and \ref{fig:gridex}(a).

To see how the procedure works, we will, for reasons of simplicity, assume a plane wave incident normally onto the surface of the mask. For a given location in the screen plane, $\mathbf{x}_\parallel'$, the phase difference, at this point on the screen, between the diffracted fields from each of the open subcells is given by the geometrical distance between each subcell and the point $\mathbf{x}_\parallel'$ on the screen. Since one in principle wants to be able to assign an arbitrary phase to the field of each cell, the phase difference over a cell should be $2\pi$ (at least). This corresponds to a path difference of $\lambda$, which is satisfied for waves propagating in the direction defined by the polar angle of transmission [see Fig.~\ref{fig:system}]
\begin{subequations}
  \label{eq:pattern_angles_of_transmission}
  \begin{align}
\theta_t\approx  \theta= \sin^{-1}\left(\frac{\lambda}{\Delta x_m}\right)
         = \sin^{-1}\left(\frac{L_s}{d}\right).
  \label{eq:pattern_angle}
\end{align}
Without loss of generality it will in the following be assumed, consistent with the illustration in Fig.~\ref{fig:system}, that the desired patterns are formed along the $x_2$-axis, i.e. the azimuthal angles of transmission  for the directions where the images are formed will be
\begin{align}
	\phi_t\approx \phi=\pm \ang{90}.
	\label{eq:pattern_angle-phi}
\end{align}
\end{subequations}
It should be noted that the angles of transmission $(\theta,\phi)$ are identical to the directions of the first order diffraction peak from a grating of period $\Delta x_m$ when illuminated at normal incidence by a wave of wavelength $\lambda$. Note also that a phase change that is a multiple of $2\pi$ can also be made to work.

Let the subgrid that is defined inside each grid cell be a $S \times S$ square lattice~\footnote{It is easy to extended the procedure to the case of a different subdivision along the two axes, but this will not be discussed here.} where $S$ denotes a positive integer [Fig.~\ref{fig:gridex}(a)]. In total $S$ different values for the phase of the field from each cell can be assigned in such a way that the minimum separation between these values are $2\pi/S$ [Fig.~\ref{fig:gridex}(a)]. When the images are formed along the $x_2$-axis, each row of the subgrid corresponds to a separate phase. Since there are $S$ subcells with the same phase, a contribution to the overall field with this phase can be chosen with a discretized amplitude of $S$ steps.

Figure~\ref{fig:gridex}(a) exemplifies the case $S=3$ which is the smallest value for $S$ that gives rise to phase differences that are not multiples of $\pi$ and leads to non-parallel vectors in the complex plane [Fig.~\ref{fig:gridex}(b)]. Moreover, Fig.~\ref{fig:gridex}(a) also presents the subcell structure for a given cell and the different relative phases associated with each row of the subcell structure. For instance, opening the central subcell seen in Fig.~\ref{fig:gridex}(a) will contribute a normalized term $\exp(\mathrm{i}4\pi/3)/3$ to the field. We have dropped the incident field and hole size from this expression since these factors will only affect the intensity and not the contrast of the final pattern. Opening one subcell from each of the three rows gives a contribution to the field that in the complex plane can be represented by the three thick blue vectors seen in Fig.~\ref{fig:gridex}(b) --- these vectors form a (hexagonal) basis (for the field) and opening more subcells will cause more steps to be taken in the directions of these vectors. Figure~\ref{fig:gridex}(b) shows as filled circles \textit{all} possible points in the complex plane reachable by different combinations of open subcells. Each row of the subgrid lattice represents a direction and a possible step in the complex plane for the field; opening $s\in\{0,1,\ldots,S\}$ subcells from row $r$ of the subgrid lattice corresponds to taking $s$ steps in the complex plane of the field along the direction vector corresponding to row $r$ [one of the thick blue vectors in Fig.~\ref{fig:gridex}(b)]. The absolute amplitude of these steps can be chosen by changing the intensity of the incident field, but only the difference between the levels is important for pattern generation.

In this way, the contribution to the mask field from a single cell can be calculated as a sum over its open subcells
\begin{align}
 \Psi_m(\mathbf{n})
  &= 
  \frac{1}{S}\sum_{r=1}^{S} 
  h\big(r|\mathbf{n}\big) \exp\left(\mathrm{i}\frac{2\pi}{S}r\right),
  \label{eq:cell-contrib}
\end{align}
where $h(r|\mathbf{n})$ is the number of open subcells in row $r$ of the cell centered at $\mathbf{x}_\parallel(\mathbf{n})$. Strictly speaking in writing  Eq.~\eqref{eq:cell-contrib} we have again neglected the incoming field. The function $h(r|\mathbf{n})$ must be chosen so that
\begin{align}
  \Psi_m(\mathbf{n}) \approx \psi_m(\mathbf{x}_\parallel(\mathbf{n})|0)
\end{align}
is satisfied with the least error; therefore, the procedure of creating a mask is reduced to selecting an optimal configuration of open subcells. This is an optimization problem, and the efficient solution will be discussed in Sec.~\ref{sec:numerical_details}. Only the relative amplitude and phase between the components of $\psi_m$ must be approximated by $\Psi_m$. The mask field can therefore be rescaled to use the approximation in $\Psi_m$ in different ways. This will also be discussed further in Sec.~\ref{sec:numerical_details}.

Any subcell from a given row of cell $\mathbf{n}$ gives an equal contribution to the field $\Psi_m(\mathbf{n})$. Therefore, it does not matter which of the subcells one opens in a row in order to give a contribution to $\Psi_m(\mathbf{n})$. This choice can therefore be made by uniform random selection over the subcells in the row.

Due to the subdivision of the cells, the discretization of the mask plane is changed, but the subcells are only used to help create an approximate field for the entire mask, so the scaling relation between the mask and the pattern in the screen plane is still given by Eq.~\eqref{eq:mask-plane-discretiatin-parameter}.
The target patterns are formed around the angles of transmissions $(\theta, \phi)$ defined in Eq.~\eqref{eq:pattern_angles_of_transmission}. The vertical coordinate of the center positions of the patterns is therefore
\begin{align}
  \label{eq:size_and_offset}
  x_2' = \pm d\tan(\theta).
\end{align}

This construction leads to a criterion for the size of the cells. Equation~\eqref{eq:pattern_angle} tells us that the method only works when the wavelength $\lambda$ is smaller than the cell width $\Delta x_m$. If the wavelength is much smaller than the cell size, the angle at which the pattern is formed will be too small and the patterns will overlap with the specular peak as well as with each other. To get good results it is therefore important to have comparable wavelength and cell size.

\section{Open-fraction optimization}

%
\begin{figure*}
  \centering
  \includegraphics[width=\textwidth]{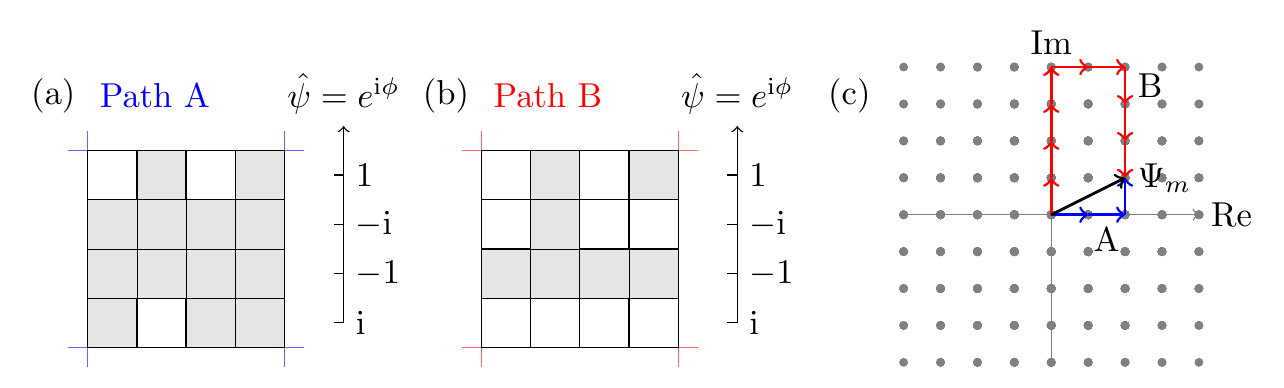}
  \caption{Illustration of the subcell-structure with $4\times 4$ subcell structure ($S=4$). The different rows have different phases $\Phi = 2\pi r/S$, with $r=1,2,3,4$. Each open subcell leads to a contribution to the overall field for the cell $\Psi_m$ in the direction of $\exp(i\Phi)$. The field from this example cell is therefore $\Psi_m = (2+\mathrm{i})/4$. Two possible paths to the desired value $\Psi_m$ is shown. The left cell takes the direct path A, while the right cell takes the longer path B, and therefore has a larger fraction of open subcells, while producing the same overall field $\Psi_m$.}
\label{fig:gridex4}
\end{figure*}

In Sec.~\ref{sec:GBBH} we presented a method that can be used to make GBHs that approximately reproduce an arbitrary pattern. The method describes which positions in a grid of possible holes that should be open and which should be closed. In a typical application of this technique a physical mask structure will be produced from this theoretical grid with open holes of a certain shape and size. The physical mask is then illuminated by an electromagnetic or a matter wave beam. Physical realizations of masks will not be discussed here, but it would be important that the holes fall on the grid positions and that their shape is similar.

Both the manufacturing step and the illumination step have certain limitations. Depending on whether the manufacturing process is done with milling or in an additive process, the ideal mask would have few or many holes. During illumination the open area of the mask should ideally be made as large as possible to minimize heating of the mask.

Because of the way that the GBH method is structured there are many different grid patterns that produce similar patterns in the screen plane. We call these masks equivalent since they produce the same or close to the same pattern. Many of the equivalent masks are the result of using a different realization of the phase $\Phi$ in Eq.~\eqref{eq:desired_screen_field} as well as the choice of which subcells to open specifically in each row. Another option for selecting between very different, but still theoretically equivalent masks is possible if the number of subdivisions $S$ for each direction of a cell is chosen to be an even number. Under this assumption, pairs of rows in the subcell structure correspond to vectors that are parallel but point in the opposite direction in the complex plane for the field [see Figs.~\ref{fig:gridex4}]. This observation allows us (for even $S$) to rewrite Eq.~\eqref{eq:cell-contrib} in the following form
\begin{align}
  \Psi_m(\mathbf{n}) 
  &= 
  \frac{1}{S}
  \sum_{r=1}^{S/2} 
  \left[ h\big(r|\mathbf{n}\big) - h\big(S/2+r|\mathbf{n}\big) \right]
    \exp\left(\mathrm{i}\frac{2\pi}{S}r\right).
    \label{eq:Psi_m_sum_rewritten}
\end{align}
Several selections for $h(r|\mathbf{n})$ can produce identical results for $\Psi_m(\mathbf{n})$. The reason is that it is no longer just the individual contributions from the rows that matter, but the difference between pairs of rows. Several different paths in the complex plane lead to exactly the same point, and therefore, to the same field $\Psi_m(\mathbf{n})$. This situation is illustrated in Fig.~\ref{fig:gridex4} where the value $S=4$ is assumed. In this figure two paths are considered --- called Path~A and Path~B --- and they do have identical 2nd and 4th rows but different 1st and 3rd rows (counted from the bottom). Even if the number of open subcells in the 1st and 3rd rows are different for the two cells, it is readily confirmed from Eq.~\eqref{eq:Psi_m_sum_rewritten} that they do produce identical contribution to the field from the $r=2$ term. This can be understood by viewing the two subcells as paths in the complex plane for the field [Fig.~\ref{fig:gridex4}(c)].

It should be noted that for the example given in Fig.~\ref{fig:gridex4}, the subcell structure of rows~\num{2} and \num{4} were deliberately set to the same for the two cells; alternatively, we could have chosen configurations for the two cells corresponding to more extreme open-to-closed subcell ratios.

If we only choose to open the minimum number of subcells in a single row in each pair of rows that produce opposite contributions, we get a minimally open mask. This is the choice made for Path A in Fig.~\ref{fig:gridex4}. If we open as many subcells as possible for each pair of rows, similar to what was done for the pair of rows~\num{1} and \num{3}, we get a maximally open solution. When changing from the minimum solution to the maximum for all the cells in a mask, the fraction of open subcells goes from $f$ to $(1-f)$. When using the minimum solution, corresponding to going directly to the desired value in the complex plane, we will at most open half of the subcells. The reason being that we want either a positive or a negative contribution from each cell and are only opening subcells in a single row for each pair of rows.

If we assume that the complex values of the desired mask field $\Psi_m$ are uniformly distributed across the available discretization area, then there is an equal amount of points that fall at every position along both the real and imaginary axes. Each of these positions on each axis is encoded with between \num{0} and \num{4} open subcells each from a total of \num{8} possible for the pair of rows. This leads to an open-fraction of \SI{27.8}{\percent} when using the minimum solution. By using the maximum method, this changes to \SI{72.2}{\percent}. These numbers are based on a completely uniform random pattern, since real masks encode patterns with correlations the minimum open-fraction will in reality be different.

It is also possible to choose an open-fraction anywhere between $f$ and $(1-f)$. This is done by not opening all the possible subcells. The cell shown in Fig.~\ref{fig:gridex4}(b) is an example of a possible intermediary choice where we have only opened the maximum amount of subcells in one of the pair of rows. We would get a similar overall field if we open more subcells in rows \num{2} and \num{4}.

\section{\label{sec:numerical_details}Numerical details}

%
\begin{figure}[tbp]
  \centering
  \includegraphics[width=0.75\columnwidth]{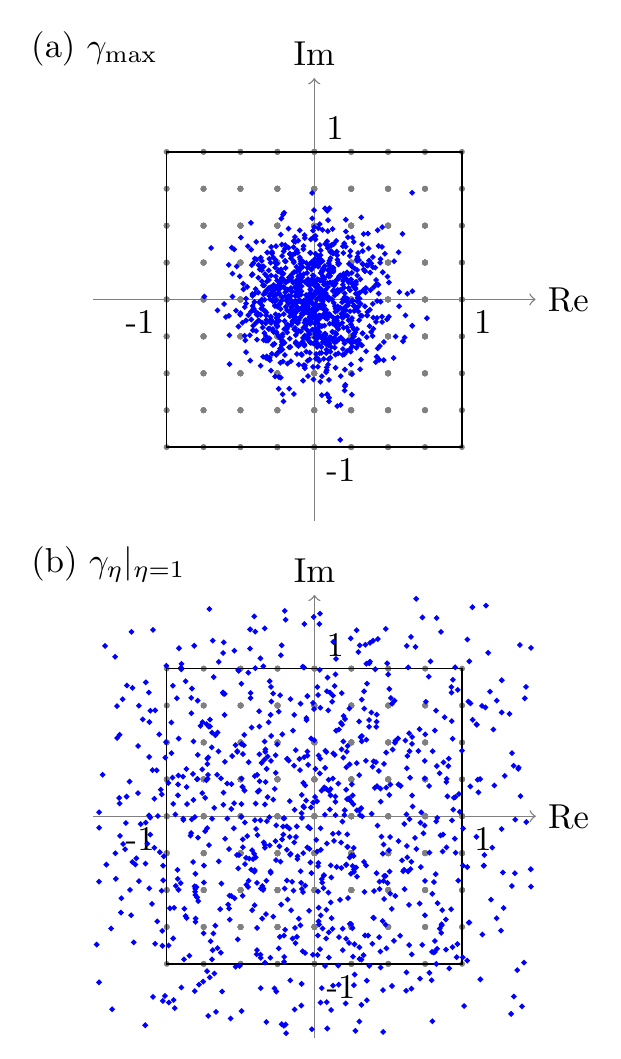}
  \caption{\label{fig:gridscale} Illustration of differently scaled mask fields. Both figures show the same \num{1000} values from the same mask field, but (a) uses $\gamma_\mathrm{max}$ for scaling and (b) uses $\gamma_\eta\rvert_{\eta=1}$. Some points fall outside the plot in (b) and have been removed.}
\end{figure}

To use the presented method for generating binary holograms we start with the pattern we want to create in the screen plane. This pattern only represents the desired intensity distribution and does not contain any phase information. We find the amplitude by taking the square root of the intensity pattern and multiply with a random phase function in accordance with Eq.~\eqref{eq:desired_screen_field} to get the field we want to construct in the screen plane. The random phase leads to a more even mask field with a more uniform amplitude, which is easier to construct with GBH. The field that we need to approximate with the hologram in each cell is then given by Eq.~\eqref{eq:discretized_mask}, which can be computed efficiently using a fast Fourier transform~\cite{Book:NumericalRecipies}.

Before the mask can be constructed, one needs to scale the amplitude of the mask field $\Psi_m(\mathbf{n})$ so that it fits within the region of the complex plane that our approximation scheme can cover, which for simplicity is defined as going from \num{-1} to \num{1} along both the real and imaginary axes in Eq.~\eqref{eq:Psi_m_sum_rewritten}. This is done by dividing the amplitude of the field by a factor $\gamma$. The rescale factor can be defined in a number of different ways. One approach is to make sure that all the values of $\Psi_m(\mathbf{n})$ actually fall within the square region available for the approximation, and this means defining the rescale factor as
\begin{equation}
	\gamma_\mathrm{max} = \max|\Psi_m(\mathbf{n})|.
	\label{eq:rescale_maximum}
\end{equation}
This rescaling sets the largest amplitude in the field to one, which means that potentially the field may go beyond the edge of the available region in the complex plane. Figure~\ref{fig:gridscale}(a) shows an example of the desired mask field for a test pattern as points in the complex plane. From Fig.~\ref{fig:gridscale}(a) we can see that the scaling option presented in Eq.~\eqref{eq:rescale_maximum} leaves parts of the available region in the complex plane uncovered by a typical pattern. Another approach to rescaling the pattern would focus on filling the available region as evenly as possible, but this will also leave some values outside the region that is available for the approximation scheme. This sort of scaling could be defined as
\begin{equation}
	\gamma_\eta = \eta\left\langle |\Psi_m(\mathbf{n})| \right\rangle_\mathbf{n},
	\label{eq:rescale_mean}
\end{equation}
where $\langle\cdot\rangle_\mathbf{n}$ defines an average over the values of the discrete mask points and $\eta$ is a parameter to tune the scaling. Figure~\ref{fig:gridscale}(b) shows a mask field scaled by $\gamma_\eta$ for $\eta=1$. When the desired mask field is rescaled in this way, the final intensity of the pattern may change, depending on, among other things, the incident field $\psi_0$ and the hole size, but the pattern will still be the same.

To perform the approximation with GBH, we need to find a choice for $h(r|\mathbf{n})$ so that the result in Eq.~\eqref{eq:cell-contrib} comes close to the required value for $\Psi_m(\mathbf{n})$. Figure~\ref{fig:gridex} illustrates the situation for $S=3$ and shows the resulting field for a cell for all choices of $h(r|\mathbf{n})$.

When the desired mask field has been rescaled to fit within the available approximation region, the selection of the appropriate $h(r|\mathbf{n})$ can be performed. This can also be done in several different ways. The simplest approach can be taken when $S=4$. In this configuration the contribution from the different rows are aligned with the axes and it is simple to take the real and imaginary part of the field and clamp them to the closest possible solution given by one or the other row pair since they correspond to two perpendicular directions in the complex plane. This will leave us with two rows that are always completely closed, and two that are potentially open. This is therefore the minimum solution described in the previous section. A maximum open-fraction is found by opening the rest of the subcells in the rows where some are already potentially open, and then opening the same number of subcells at random in the corresponding row. To find a mask with a specific open-fraction, one approach is to pairwise open or close subcells at random in pairs of rows within a randomly selected cell until the desired open-fraction is reached.

For an arbitrary value of $S$ a way to find the optimal approximation is to precompute $\Psi_m(\mathbf{n})$ for a cell using Eq.~\eqref{eq:cell-contrib} given all possible combinations of choices for $h(r|\mathbf{n})$ and store the results in a search-tree. We can then search through all the options and pick the closest approximation for each cell. The number of possible options grows very rapidly with $S$. This procedure leads to the number of open subcells that are required for each row in each individual cell of the mask. Which subcells to open can then be selected at random.

These methods leave us with one possible realization of a mask for a given target pattern. This grid structure tells us where we need holes on a square grid to create a GBH mask for our pattern. The mask can be manufactured physically by creating an array of holes with similar geometry in a film that is suitable to block our incident wave. It is also possible to check the masks by performing simulations of the effect of the mask. This can either be done through a direct calculation based on Eq.~\eqref{eq:fraunhofer} or by performing a more physically accurate simulation that takes into account the final mask geometry. The direct calculation is convenient since it is mainly a Fourier transform of the mask structure that can be performed using an FFT. The physical mask geometry can be taken into account by either performing the Fourier transform on a large discretized version of the mask geometry or by taking the superposition of analytic diffraction solutions for the individual holes in the mask.

\section{Examples}

%
\begin{figure*}[tbp]
  \centering
  \includegraphics[width=\textwidth]{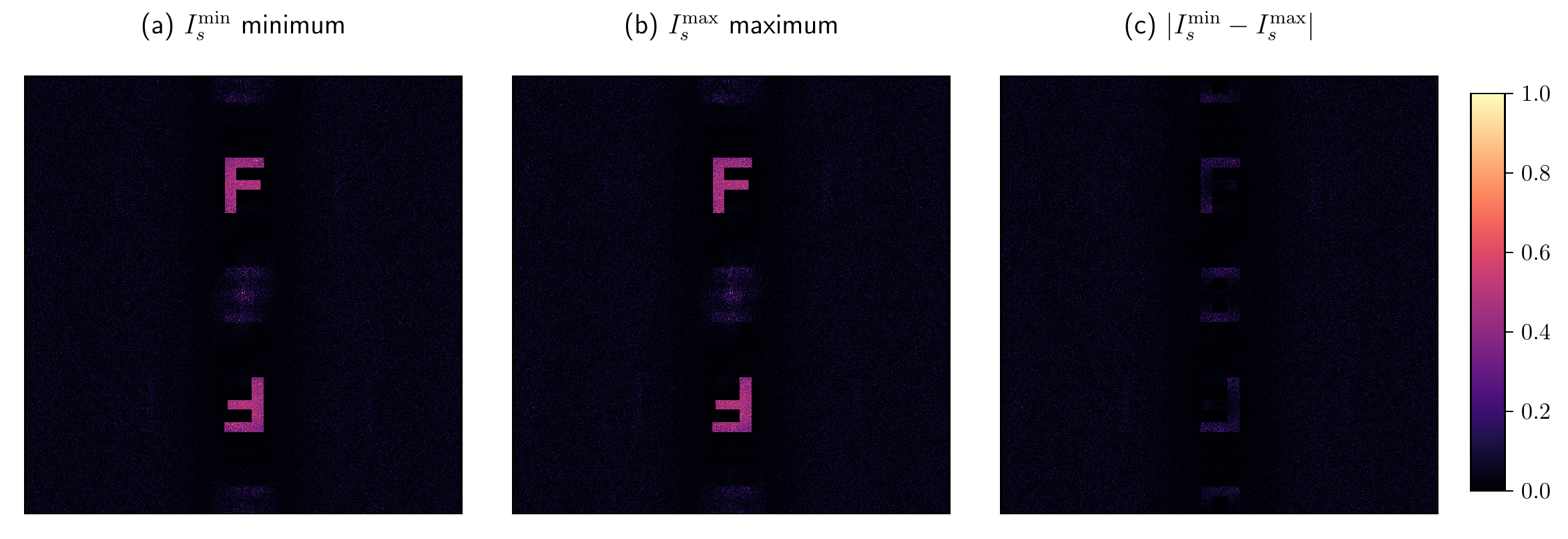}
  \caption{\label{fig:F_min_max_diff}Simulation results for masks created for the same target pattern. The $I^\mathrm{min}_s$ figure shows the result from a minimally opened mask, while the $I^\mathrm{max}_s$ shows the result from a maximally opened mask. The difference figure shows the difference $|I^\mathrm{min}_s - I^\mathrm{max}_s|$ between the two versions. All axes are similar and the intensity values are comparable and of arbitrary unit. Parts of the specular peak for $I^\mathrm{max}_s$ is saturated above the scale.}
\end{figure*}

Examples will now be given of different GBH masks generated by the method described in the preceeding section, and their performance will be evaluated by simulation. First we will look qualitatively at how we can adjust the open-fraction of the mask, and see what impact this has on the generated pattern. Then we will perform a more quantitative investigation, where we investigate the behavior of both the contrast and the error tolerance of the masks as we force different open-fractions. Finally we will study the contrast and the error tolerance of masks generated with different scaling of the mask fields.

The simulations presented in the following sections were all performed for binary masks generated with subdivision $S=4$ since this allows us to generate the mask without large lookup tables and it is the smallest subdivision that allows for easy adjustment of the open-fraction of a mask by opening and closing hole pairs. The intensity patterns created by the masks were found using the direct method of Fraunhofer propagation described by Eq.~\eqref{eq:fraunhofer}. When nothing else is specified the mask patterns were rescaled using the maximum scale factor $\gamma_\mathrm{max}$ given by Eq.~\eqref{eq:rescale_maximum}, while $\gamma_\eta$ was used in cases where $\eta$ is given.

%
\begin{figure}[tbp]
  \centering
  \includegraphics[width=\columnwidth]{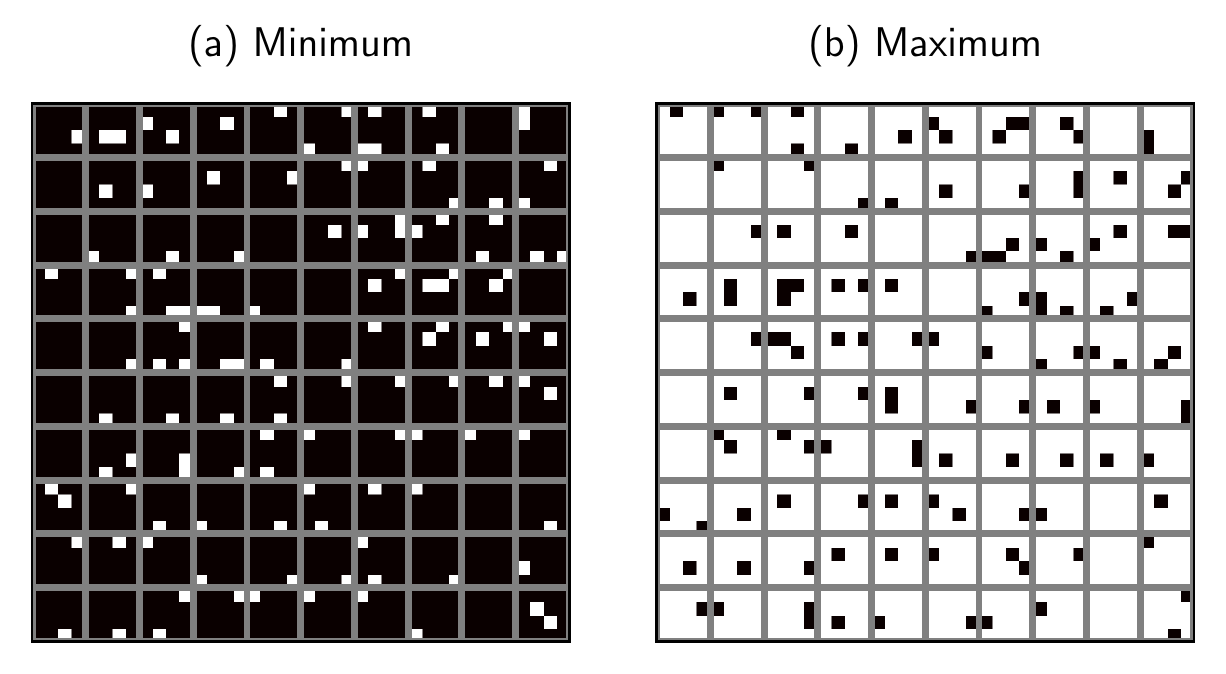}
    \caption{\label{fig:sbs-masks}Parts of the masks (size $40 \times 40$ subcells) used to generate the two intensity patterns presented in Fig.~\ref{fig:F_min_max_diff}: (a) mask with the minimum number of holes (open subcells); (b) mask with the maximum number of holes.  Light (dark) colored squares represent open (closed) subcells. Gray lines show cell boundaries.}
\end{figure}

Figure~\ref{fig:F_min_max_diff} presents the central area of two simulated intensity patterns in the screen plane. The masks used to obtain the patterns in the screen plane depicted in Figs.~\ref{fig:F_min_max_diff}(a) and \ref{fig:F_min_max_diff}(b) were generated by assuming the same target pattern; an image of the character \textit{F}. The width and height of the character were approximately a third of the width and height of the pattern area. The image used for the target had a resolution of \num{600}$\times$\num{600} pixels. When using a scheme with $S=4$ subdivisions, this leads to a mask with \num{2400}$\times$\num{2400} subcells. For the $I^\text{min}_s$-pattern the mask was created with the minimum number of open subcells, while for the $I^\text{max}_s$-pattern the maximum number of subcells were opened. Figure~\ref{fig:sbs-masks} shows cutouts of the same corner of the two masks used to generate the results in Figs.~\ref{fig:F_min_max_diff}(a) and \ref{fig:F_min_max_diff}(b). The minimally open mask had in this case \SI{7.3}{\percent} open subcells, while the maximally open mask had \SI{92.7}{\percent} open subcells. These numbers are not necessarily the same for other mask realizations because of the random phase that is applied in the beginning. These two masks were produced using the same random phase for the target field. The number of the open subcells is inverted when going from the minimum to the maximum solution, but the masks themselves are not necessarily inverted. A closer inspection of the two cutouts in Fig.~\ref{fig:sbs-masks} makes it apparent that the number of open subcells in a cell is inverted, but which subcells that are chosen to be opened or closed in a particular row is picked at random.

Figures~\ref{fig:F_min_max_diff}(a) and \ref{fig:F_min_max_diff}(b) show the intensity distribution behind the mask with the minimum and the maximum number of open holes respectively. The resulting intensity patterns $I^\text{min}_s$ and $I^\text{max}_s$ above and below the position of the center specular peak are almost identical. The absolute difference between them is presented in Fig.~\ref{fig:F_min_max_diff}(c). The main difference is a speckle pattern around the borders of the pattern. The speckle noise is higher towards the edges than at the center of the patterns. This noise depends on the random selections made when constructing the two different masks. This leads to visible noise in the difference image shown in Fig.~\ref{fig:F_min_max_diff}(c). This effect is most pronounced along the vertical axis, with noise primarily at the top and bottom of the pattern. The reason is most likely that this is the direction away from the center maxima in which the pattern is formed, and that the variation away from the path length assumed in the approximations is at its largest. If the mask is directly inverted when going from a minimum to maximum solution, so that the random choices are similar, the intensity in the patterns become identical with the same noise.

The results shown in Figs.~\ref{fig:F_min_max_diff}(a) and \ref{fig:F_min_max_diff}(b) are comparable and the colorbar is adjusted to show the strongest intensity observed in the pattern area of either of the two realizations. The specular point at the center is much stronger than the rest of the figure and oversaturates the colorbar. Opening a larger area of the mask does not affect the patterns that are generated, but the extra intensity goes into the specular point. This point is a single pixel in these simulations, and the intensity value of this pixel changes by several orders of magnitude when going from the minimum to the maximum solution. If the simulations take the actual geometry of the holes into account, the central peak would broaden due to diffraction and there would be a visible difference between the central area of the two patterns.

%
\begin{figure*}[tbp]
  \centering
  \includegraphics[width=\textwidth]{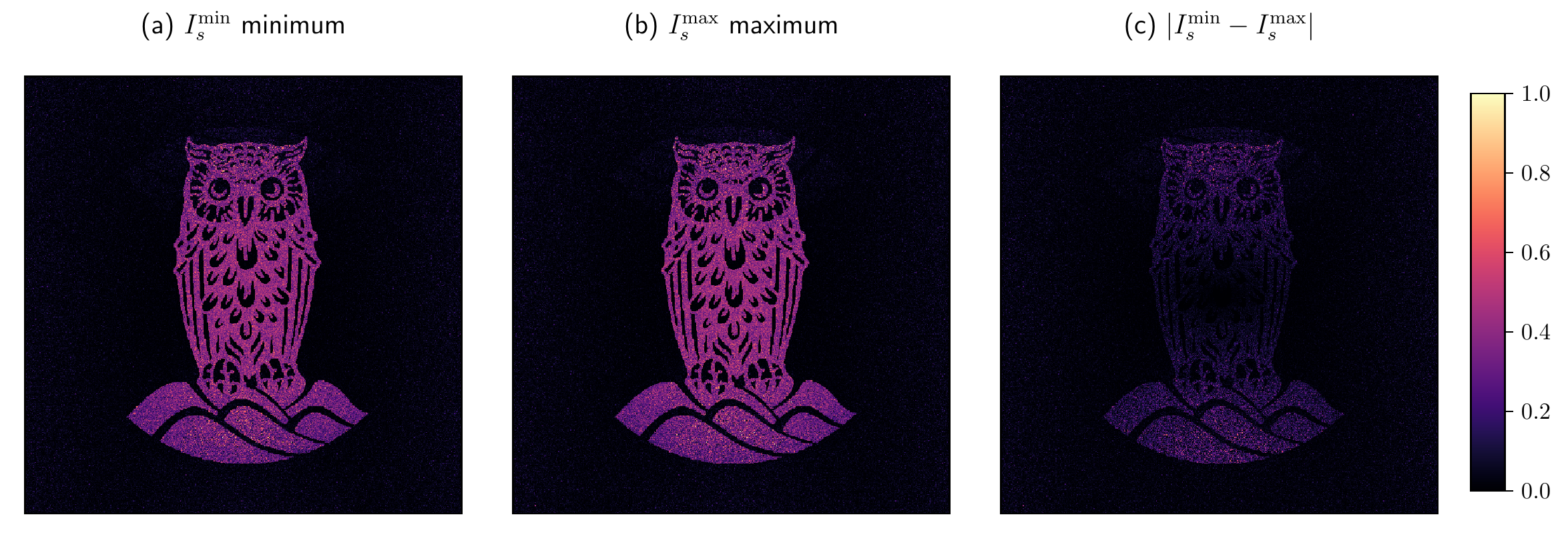}
  \caption{\label{fig:owl_min_max_diff}Similar to Fig.~\ref{fig:F_min_max_diff}, with more complex test pattern.  Shows only the intensity in the target area where the pattern is formed.}
\end{figure*}

Figure~\ref{fig:owl_min_max_diff} shows results similar to those presented in Fig.~\ref{fig:F_min_max_diff}, but for a much more complex pattern of an owl. The target image still had a resolution of \num{600}$\times$\num{600} pixels. In this case the minimal mask had $7.6\%$ open subcells, while the maximum mask had $92.4\%$. As with the intensity patterns in Fig.~\ref{fig:F_min_max_diff} the difference between the minimum and maximum pattern is primarily in the noise in the edges of the pattern [Fig.~\ref{fig:owl_min_max_diff}(c)] due to the randomly chosen phase. Note that the images in Fig.~\ref{fig:owl_min_max_diff} are cutouts of the full simulation results showing only the region where the pattern is formed.

\section{Contrast measurements}
According to Eq.~\eqref{eq:Psi_m_sum_rewritten}, the pattern formed from masks with an even number of subcell rows (i.e. value of $S$) will only depend on the difference between the number of open subcells in the row pairs. This is an approximation, and to investigate how well it holds we need to compare the patterns generated from several equivalent masks with different open-fractions. The goal of the GBH method is to create sharp, arbitrary patterns. How well different masks are able to perform this task can be quantified by measuring the contrast of the final pattern. We define the contrast as
\begin{equation}
\alpha = \frac{|\bar{I}_\text{in}-\bar{I}_\text{out}|}{|\bar{I}_\text{in}+\bar{I}_\text{out}|},
\end{equation}
where $\bar{I}_\text{in}$ is the average intensity inside our pattern and $\bar{I}_\text{out}$ is the average intensity outside [see Fig.~\ref{fig:target}]. We only look at the intensity that falls inside the area predicted by Eq.~\eqref{eq:size_and_offset}, and regard the parts of the pattern area we want to expose as inside our pattern and the other parts as outside. This definition only works for binary patterns with no intermediary values.

%
\begin{figure}[htbp]
  \centering
  \includegraphics[width=\columnwidth]{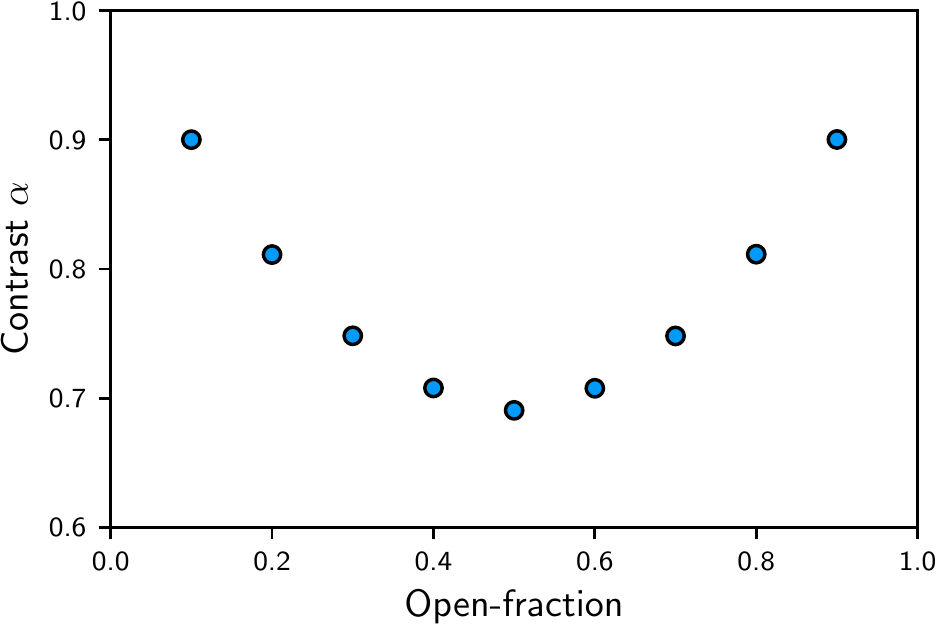}
  \caption{\label{fig:alpha_vs_openpart_sbs_a}Contrast $\alpha$ of test patterns as a function of the open-fraction of the masks used when making the patterns. Average intensities are based on \num{10} mask realizations.}
\end{figure}

%
\begin{figure}[htbp]
  \centering
  \includegraphics[width=\columnwidth]{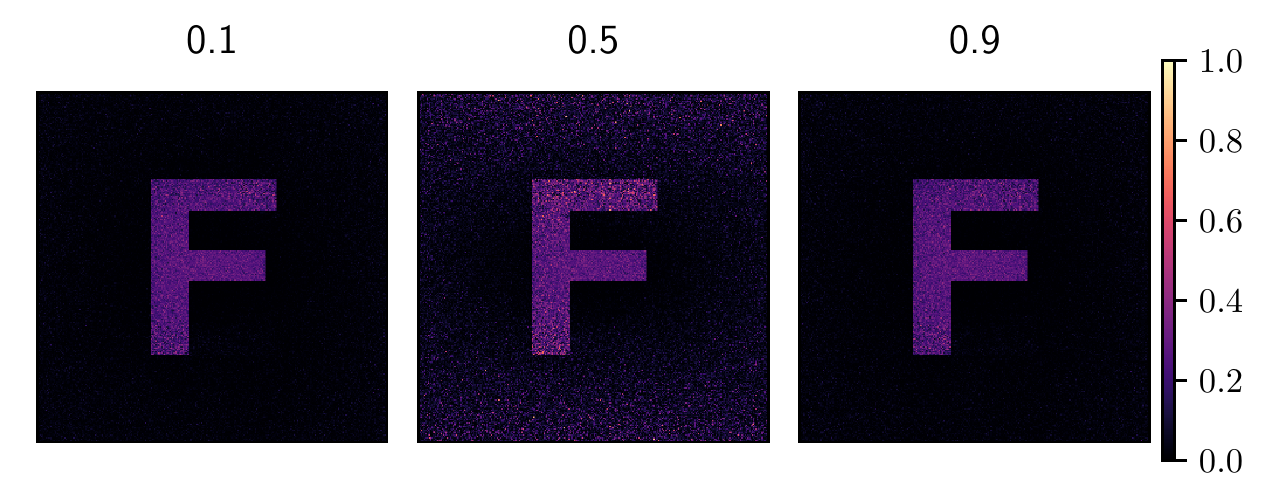}
  \caption{\label{fig:alpha_vs_openpart_sbs_b}Intensity patterns created from masks of open-fractions 0.1, 0.5 and 0.9 (left-to-right.}
\end{figure}

Several different masks corresponding to the F-pattern shown in Fig.~\ref{fig:target} were generated for different open-fractions in the range from \SI{10}{\percent} to \SI{90}{\percent}. The resulting contrast measurements $\alpha$ are shown in Fig.~\ref{fig:alpha_vs_openpart_sbs_a} as a function of the open-fraction. The mean intensities were found by looking at the intensity patterns generated by \num{10} different masks for each open-fraction level. These masks were generated using the same scaling $\gamma_\mathrm{max}$, but with different random phases. Figure~\ref{fig:alpha_vs_openpart_sbs_a} shows that the contrast is highest for masks close to the minimum or maximum number of open subcells and goes down towards an open-fraction of \SI{50}{\percent}. Figure \ref{fig:alpha_vs_openpart_sbs_b} shows three intensity patterns created by masks of open-fraction \SI{10}{\percent}, \SI{50}{\percent} and \SI{90}{\percent}, respectively. When the open-fraction is around \SI{50}{\percent}, there is a clear increase in the mean value for the intensity outside of the pattern. Similar behavior is also observed for other patterns, and when using more complex methods for calculating the intensity pattern that takes the hole geometry into account.

Inspection of Fig.~\ref{fig:alpha_vs_openpart_sbs_b} and several other similar test patterns show that there is an increase in the intensity of the noise around the edges of the target area when we increase the open-fraction. Test patterns where a large part of the pattern occupies the area next to the edges behave differently to what is shown in Fig.~\ref{fig:alpha_vs_openpart_sbs_a}. The reason is that there is a much larger amount of noise in the inside area when it is located along the edge.

The approximations that are used when generating the masks hold better close to the center of the pattern area. This could explain why the intensity patterns in general become more noisy in regions away from the center. For masks with an open-fraction of \SI{50}{\percent} a lot of extra holes have been added to the pattern. According to the approximations these should cancel perfectly, but in practice this is not the case. For the masks with minimum open-fraction, all of the open subcells directly contribute to forming the pattern. For masks that have \SI{50}{\percent} open subcells, extra pairs that might not cancel completely have been added, resulting in a higher amount of noise. For open-fraction above \SI{50}{\percent}, the contrast improves again as the mask moves towards what is effectively an inverted mask from the minimum configuration [see Fig.~\ref{fig:alpha_vs_openpart_sbs_a}]. For masks with the maximum open-fraction, the cancellations work better since there is now a much higher number of open subcells.

\subsection{Tolerance to errors}

%
\begin{figure}[htbp]
  \centering
  \includegraphics[width=\columnwidth]{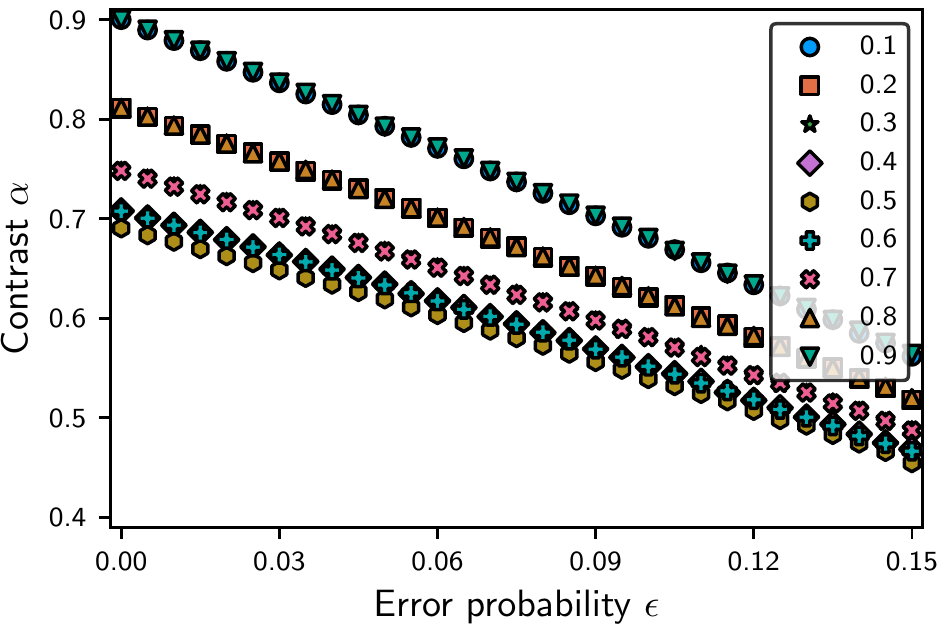}
  \caption{\label{fig:alpha_vs_err}Contrast $\alpha$ as a function of the probability $\epsilon$ of having an error in a subcell. The different symbols correspond to masks created with several different initial open-fractions. Series with complementing open-fraction ($f$ and $1-f$) are overlapping. Each data point is based on the mean contrast $\alpha$ from intensity patterns computed for a series of mask realisations.}
\end{figure}

We now turn to the discussion of error tolerance of the generated masks. If we introduce errors in the masks by randomly changing open and closed subcells with a certain error probability, we can use the change in the contrast as a measure for the robustness of the mask. Figure \ref{fig:alpha_vs_err} shows the robustness for masks created with several different open-fractions. As in Fig.~\ref{fig:alpha_vs_openpart_sbs_a} the behavior is symmetric around an open-fraction of $50\%$ with pairs of overlapping curves in the figure. As the error probability increases the contrast goes down for all the different masks, which is the expected behavior. When we introduce errors we change the mask towards a mask with a random hole pattern on a grid. This means that the contrast should not go towards zero, but would converge towards the value associated with the response of a random mask.

\begin{figure}[htbp]
  \centering
  \includegraphics[width=\columnwidth]{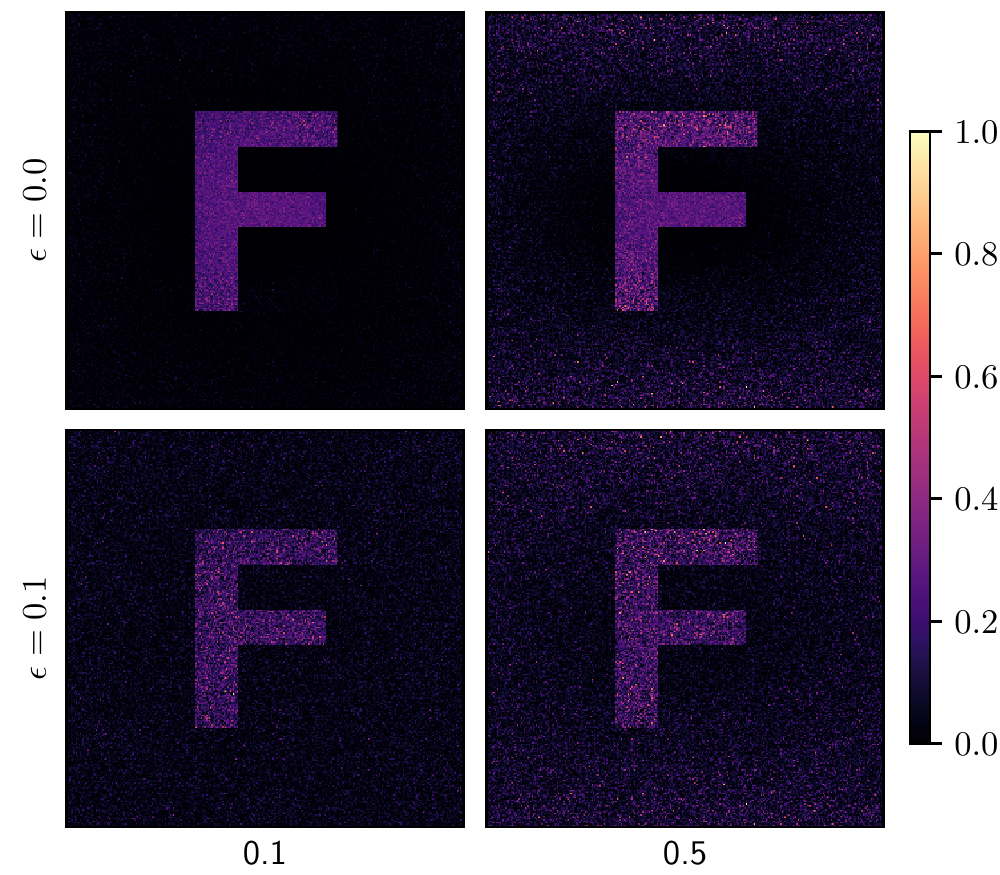}
  \caption{\label{fig:F_sbs} Simulated response from masks made with open-fractions of \num{0.1} (first column) and \num{0.5} (second column) and error probabilities \num{0.0} (first row) and~\num{0.1} (second row). All other parameters were held constant when generating the different masks. The intensity values are comparable and of arbitrary unit.}
\end{figure}

We see from Fig.~\ref{fig:alpha_vs_err} that the spread in robustness goes down as we increase the error probability. The $10\%$ and $90\%$ masks start with the highest contrast, but also decreases faster than any of the other open-fractions in between. The contrast of the pattern  corresponding to the mask created with $50\%$ open subcells is more robust as errors are introduced. Its contrast falls off slower than what is observed for all other open-fractions. The reason for this is probably that it started with a high amount of noise. It has the worst contrast of all the choices of open-fraction made. The behavior observed in Fig.~\ref{fig:alpha_vs_err} continues if we raise the error probability above $0.15$, with the contrast for the other open-fractions converging towards the contrast of the \SI{50}{\percent} open case. Figure~\ref{fig:F_sbs} presents a comparison of patterns generated using different masks. The left column and the right column show the response from masks with \SI{10}{\percent} and \SI{50}{\percent} open subcells respectively. The two rows shows masks with error probability $\epsilon=0.0$ and $0.1$.

%

\section{Scaling mask pattern}

%
\begin{figure}[tbp]
  \centering
  \includegraphics[width=\columnwidth]{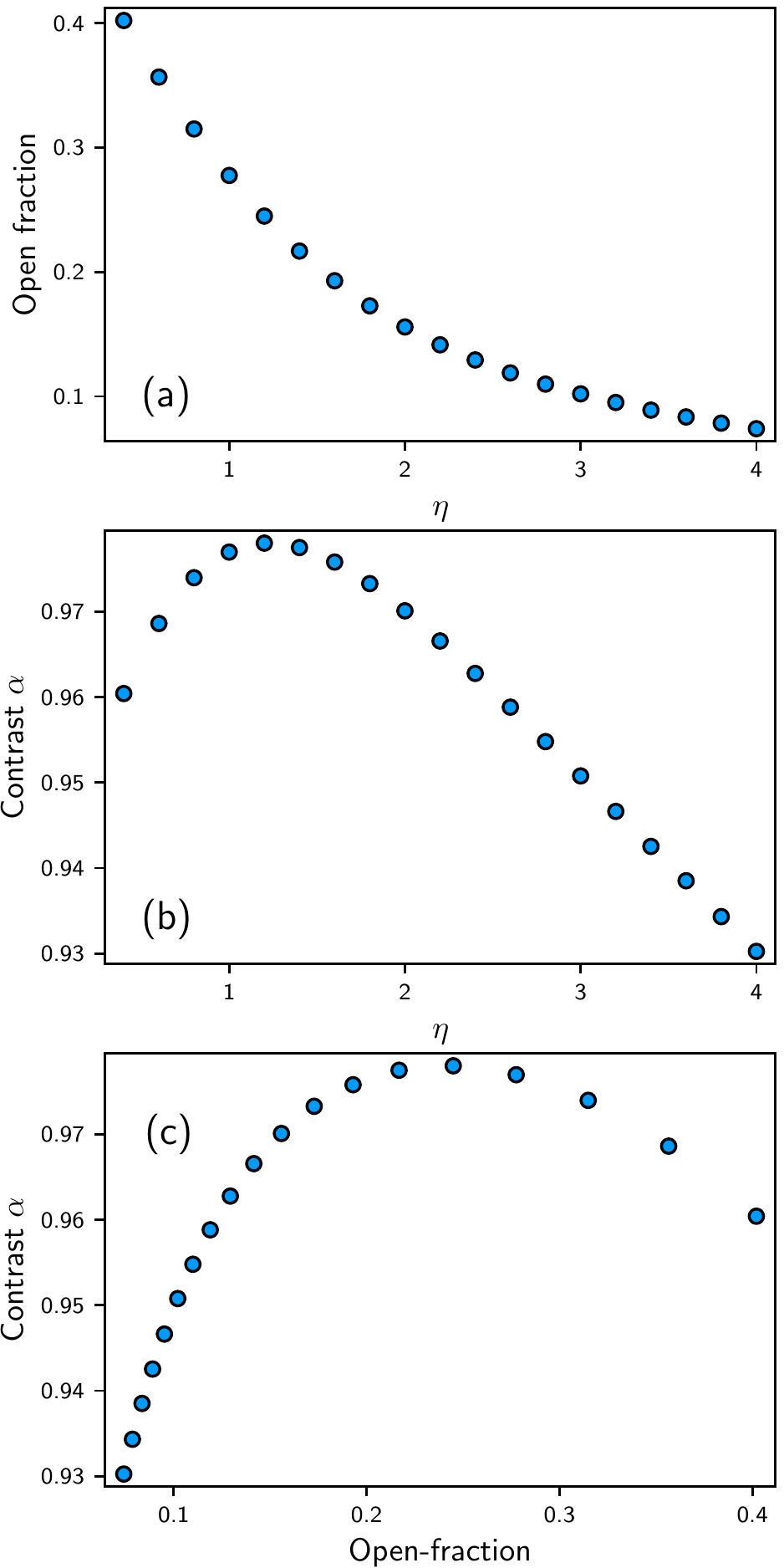}
  \caption{\label{fig:eta_of_alpha} Using the minimum method for generating masks with different field scaling $\gamma_\eta$ shows a variation in (a) open-fraction of the subcells and (b) contrast $\alpha$ as a function of the scaling parameter $\eta$. (c) Contrast $\alpha$ as a function of the open-fraction controlled by change in the mask field scaling $\gamma_\eta$.}
\end{figure}

We will now investigate the dependence of the contrast and the tolerance to error for masks created from an initial field that is scaled in such a way that not all possible values fit within the range that is representable by the discretized GBH masks [see Eq.~\eqref{eq:rescale_mean}]. To this end, we will use a procedure very similar to what was used in the previous section, for instance, to generate the results of Fig.~\ref{fig:F_min_max_diff}. The only difference is that the rescale step is now performed using $\gamma_\eta$ given by Eq.~\eqref{eq:rescale_mean} for a range of $\eta$. When $\eta=4$ the new rescaling factor is approximately equal to the old one, $\gamma_\eta \approx \gamma_\text{max}$. All of the results presented here were generated using the minimum method.

Figure~\ref{fig:eta_of_alpha} presents both the open-fraction of subcells and the contrast for a series of different scaling parameters $\eta$. Figure~\ref{fig:eta_of_alpha}(a) shows that for a smaller scaling parameter $\eta$ the open-fraction of the mask is larger. The reason is that a larger fraction of the components of the mask field lie in the outer part or outside the discretization region with a smaller $\eta$. At the same time there is also a change in the measured contrast for the patterns created by these masks, as shown in Fig.~\ref{fig:eta_of_alpha}(b). The variation in the contrast is smaller than the variation observed in the previous section, but the results show a smooth behavior and a maximum contrast for $\eta = 1.2$. When the value of $\eta$ is taken to be smaller than \num{4} the rescale value becomes larger and the components of the mask field better fill out the discretization region. This leads to a more even use of the available levels and a higher degree of accuracy when representing a large part of the field values, but also a limited scale that does not have the range to accurately represent the larger amplitudes. As we decrease $\eta$ from \num{4}, the contrast initially improves as the mask pattern is better encoded. When $\eta < 1$, points with amplitude equal to the mean are shifted to the top of the encodable region, a larger part of the mask values are clamped to the size of the region and not represented properly, and the contrast of the produced pattern drops again.

Figure~\ref{fig:eta_of_alpha}(c) removes the direct dependence on $\eta$ by combining the results of Figs.~\ref{fig:eta_of_alpha}(a) and \ref{fig:eta_of_alpha}(b) and creates a comparable plot to Fig.~\ref{fig:alpha_vs_openpart_sbs_a}. This figure shows a very different behavior from what we saw previously. Now the change in open hole fraction is connected to a change in how the approximation scheme is utilized. The change in contrast isn't as large as we saw in the previous sections, but there is a clear trend which shows a maximum contrast at an open part fraction close to $0.25$. This scaling gives a very uniform distribution of mask field points in the approximation region, and thus represents a solution that comes close to the best utilization of the approximation scheme.

Comparing the results of Fig.~\ref{fig:alpha_vs_openpart_sbs_a} to Fig.~\ref{fig:eta_of_alpha}(c) shows that the better way of changing the open-fraction of subcells in a mask is done by changing the way the discretization scheme in the GBH approximation is used. This allows for a change of open-fraction in almost the entire range between \num{0.0} and \num{0.5}, while still keeping a similar or better contrast than the original solution investigated using $\gamma_\text{max}$, as opposed to the scheme of adding additional open subcells by using the different solutions to Eq.~\eqref{eq:Psi_m_sum_rewritten}. Additional holes can still be added to change the open-fraction and therefore extend this scheme to work for open-fractions above \num{0.5}. These results are not necessarily similar for all target patterns. Different patterns leads to mask fields that are differently distributed in the complex plane. The field scaling that produces the most uniform utilization of the available approximation region will therefore depend on the pattern. However, several mask patterns were investigated, and they all showed similar behavior. Optimal contrast for a given pattern and desired open fraction was found by scanning over the parameter space of interest. A more sophisticated optimization technique could also have been used, but we did not explore this.

%
\begin{figure}[tbp]
  \centering
  \includegraphics[width=\columnwidth]{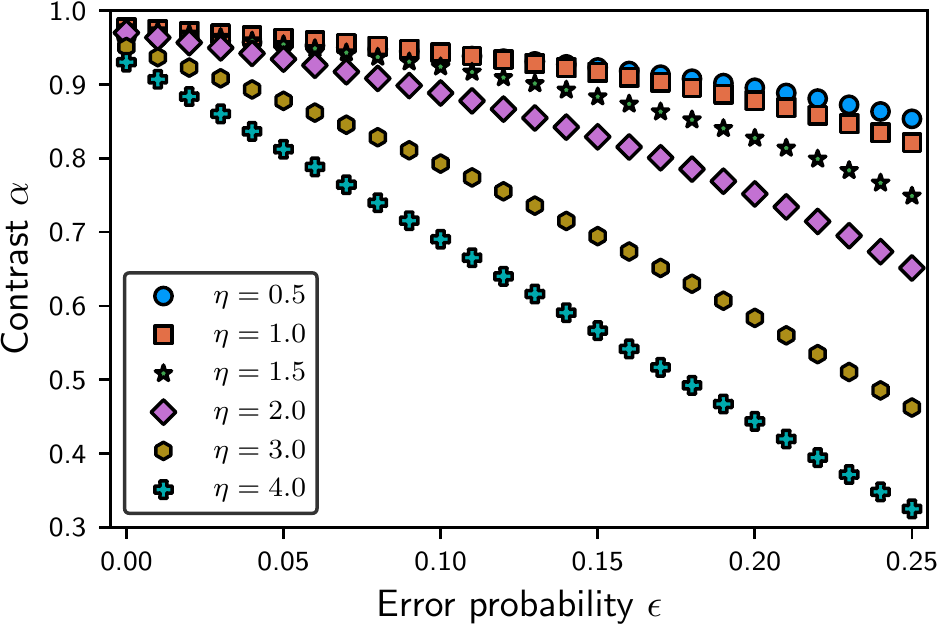}
  \caption{\label{fig:alpha_err_eta} Contrast $\alpha$ as a function of the probability of having an error in a subcell of the mask for masks created different scaling $\eta$. Each data point is based on the mean contrast $\alpha$ from intensity patterns computed for a series of mask realisations.}
\end{figure}

Figure~\ref{fig:alpha_err_eta} presents the robustness of masks generated with different $\gamma_\eta$. This figure is similar to Fig.~\ref{fig:alpha_vs_err}, but shows the contrast behavior up to the higher error probability of $0.25$. The series with $\eta=4.0$ is close to the best cases presented previously using the maximum scaling, this is as expected since $\gamma_\text{max} \approx \gamma_{\eta}|_{\eta=4.0}$ for this test pattern. The series that correspond to using a scaling parameter $\eta$ that better utilize the available approximation region perform better. For series with $\eta < 1.5$ the contrast is above $0.9$ when \SI{10}{\percent} of the subcells are flipped (i.e. $\epsilon=0.1$). These values for $\eta$ correspond to masks with an open-fraction from just under \num{0.2} and up to almost \num{0.5}. This high level of robustness shows that using more subcells to encode the signal doesn't only improve the initial contrast, it also improves the robustness of the masks to errors in which subcells are opened or closed at random.

\section{Conclusion and future work}
We have presented a new technique that extends the grid based binary holography method by Onoe and Kaneko~\cite{Onoe79}. The new technique lets us freely decide between grids with a low or high number of open subcells. This allows the grid based holography method to be adjusted to better fit specific user needs, whether this means tuning the open hole fraction down for faster production and more rigid masks, or tuning it up to prevent problems such as heating of the mask during exposure. The technique can be extended to work for any selected open-fraction between the two extreme solutions shown, and investigations into the contrast of the patterns produced with different masks lead us to the conclusion that choosing an intermediary open-fraction between the extrema is best done by scaling the initial field. This scaling impacts both the contrast of the generated patterns and the open-fraction of the generated masks, leading to a tradeoff between these two parameters when selecting the scaling for generating masks. All results presented here work in the Fraunhofer diffraction limit. Future work should include the integration of lenses to reduce the size of the optical system and improve resolution. 

\begin{acknowledgments}
The authors gratefully acknowledge support from the Research Council of Norway, Fripro Project 213453 and Forny Project 234159. The research of I.S. was supported in part by the Research Council of Norway Contract No. 216699 and The French National Research Agency~(ANR) under contract ANR-15-CHIN-0003-01.
\end{acknowledgments}

\nocite{apsrev41Control}
\bibliographystyle{apsrev4-1}
\bibliography{bibtex.bib}

\begin{thebibliography}{28}%
\makeatletter
\providecommand \@ifxundefined [1]{%
 \@ifx{#1\undefined}
}%
\providecommand \@ifnum [1]{%
 \ifnum #1\expandafter \@firstoftwo
 \else \expandafter \@secondoftwo
 \fi
}%
\providecommand \@ifx [1]{%
 \ifx #1\expandafter \@firstoftwo
 \else \expandafter \@secondoftwo
 \fi
}%
\providecommand \natexlab [1]{#1}%
\providecommand \enquote  [1]{``#1''}%
\providecommand \bibnamefont  [1]{#1}%
\providecommand \bibfnamefont [1]{#1}%
\providecommand \citenamefont [1]{#1}%
\providecommand \href@noop [0]{\@secondoftwo}%
\providecommand \href [0]{\begingroup \@sanitize@url \@href}%
\providecommand \@href[1]{\@@startlink{#1}\@@href}%
\providecommand \@@href[1]{\endgroup#1\@@endlink}%
\providecommand \@sanitize@url [0]{\catcode `\\12\catcode `\$12\catcode
  `\&12\catcode `\#12\catcode `\^12\catcode `\_12\catcode `\%12\relax}%
\providecommand \@@startlink[1]{}%
\providecommand \@@endlink[0]{}%
\providecommand \url  [0]{\begingroup\@sanitize@url \@url }%
\providecommand \@url [1]{\endgroup\@href {#1}{\urlprefix }}%
\providecommand \urlprefix  [0]{URL }%
\providecommand \Eprint [0]{\href }%
\providecommand \doibase [0]{http://dx.doi.org/}%
\providecommand \selectlanguage [0]{\@gobble}%
\providecommand \bibinfo  [0]{\@secondoftwo}%
\providecommand \bibfield  [0]{\@secondoftwo}%
\providecommand \translation [1]{[#1]}%
\providecommand \BibitemOpen [0]{}%
\providecommand \bibitemStop [0]{}%
\providecommand \bibitemNoStop [0]{.\EOS\space}%
\providecommand \EOS [0]{\spacefactor3000\relax}%
\providecommand \BibitemShut  [1]{\csname bibitem#1\endcsname}%
\let\auto@bib@innerbib\@empty
\bibitem [{\citenamefont {Borkar}\ and\ \citenamefont
  {Chien}(2011)}]{Borkar11}%
  \BibitemOpen
  \bibfield  {author} {\bibinfo {author} {\bibfnamefont {S.}~\bibnamefont
  {Borkar}}\ and\ \bibinfo {author} {\bibfnamefont {A.~A.}\ \bibnamefont
  {Chien}},\ }\bibfield  {title} {\enquote {\bibinfo {title} {{The Future of
  Microprocessors}},}\ }\href@noop {} {\bibfield  {journal} {\bibinfo
  {journal} {Commun. ACM}\ }\textbf {\bibinfo {volume} {54}},\ \bibinfo {pages}
  {67} (\bibinfo {year} {2011})}\BibitemShut {NoStop}%
\bibitem [{ITR(2015)}]{ITRS}%
  \BibitemOpen
  \href {http://www.itrs2.org} {\emph {\bibinfo {title} {ITRS 2.0:
  International Technology Roadmap for Semiconductors, More Moore}}},\ \bibinfo
  {type} {Tech. Rep.}\ (\bibinfo {year} {2015})\BibitemShut {NoStop}%
\bibitem [{\citenamefont {Berggren}\ \emph {et~al.}(1995)\citenamefont
  {Berggren}, \citenamefont {Bard}, \citenamefont {Wilbur}, \citenamefont
  {Gillaspy}, \citenamefont {Helg}, \citenamefont {McClelland}, \citenamefont
  {Rolston}, \citenamefont {Phillips}, \citenamefont {Prentiss},\ and\
  \citenamefont {Whitesides}}]{Berggren95}%
  \BibitemOpen
  \bibfield  {author} {\bibinfo {author} {\bibfnamefont {K.~K.}\ \bibnamefont
  {Berggren}}, \bibinfo {author} {\bibfnamefont {A.}~\bibnamefont {Bard}},
  \bibinfo {author} {\bibfnamefont {J.~L.}\ \bibnamefont {Wilbur}}, \bibinfo
  {author} {\bibfnamefont {J.~D.}\ \bibnamefont {Gillaspy}}, \bibinfo {author}
  {\bibfnamefont {A.~G.}\ \bibnamefont {Helg}}, \bibinfo {author}
  {\bibfnamefont {J.~J.}\ \bibnamefont {McClelland}}, \bibinfo {author}
  {\bibfnamefont {S.~L.}\ \bibnamefont {Rolston}}, \bibinfo {author}
  {\bibfnamefont {W.~D.}\ \bibnamefont {Phillips}}, \bibinfo {author}
  {\bibfnamefont {M.}~\bibnamefont {Prentiss}}, \ and\ \bibinfo {author}
  {\bibfnamefont {G.~M.}\ \bibnamefont {Whitesides}},\ }\bibfield  {title}
  {\enquote {\bibinfo {title} {Microlithography by using neutral metastable
  atoms and self-assembled monolayers},}\ }\href {\doibase
  10.1126/science.7652572} {\bibfield  {journal} {\bibinfo  {journal}
  {Science}\ }\textbf {\bibinfo {volume} {269}},\ \bibinfo {pages} {1255}
  (\bibinfo {year} {1995})}\BibitemShut {NoStop}%
\bibitem [{\citenamefont {Baldwin}(2005)}]{Baldwin05}%
  \BibitemOpen
  \bibfield  {author} {\bibinfo {author} {\bibfnamefont {K.}~\bibnamefont
  {Baldwin}},\ }\bibfield  {title} {\enquote {\bibinfo {title} {Metastable
  helium: Atom optics with nano-grenades},}\ }\href {\doibase
  10.1080/00107510412331332798} {\bibfield  {journal} {\bibinfo  {journal}
  {Contemp. Phys.}\ }\textbf {\bibinfo {volume} {46}},\ \bibinfo {pages} {105}
  (\bibinfo {year} {2005})}\BibitemShut {NoStop}%
\bibitem [{\citenamefont {Lohmann}\ and\ \citenamefont
  {Paris}(1967)}]{Lohmann67}%
  \BibitemOpen
  \bibfield  {author} {\bibinfo {author} {\bibfnamefont {A.~W.}\ \bibnamefont
  {Lohmann}}\ and\ \bibinfo {author} {\bibfnamefont {D.~P.}\ \bibnamefont
  {Paris}},\ }\bibfield  {title} {\enquote {\bibinfo {title} {{Binary
  Fraunhofer Holograms, Generated by Computer}},}\ }\href {\doibase
  10.1364/AO.6.001739} {\bibfield  {journal} {\bibinfo  {journal} {Appl. Opt.}\
  }\textbf {\bibinfo {volume} {6}},\ \bibinfo {pages} {1739} (\bibinfo {year}
  {1967})}\BibitemShut {NoStop}%
\bibitem [{\citenamefont {{Onoe}}\ and\ \citenamefont
  {{Kaneko}}(1979)}]{Onoe79}%
  \BibitemOpen
  \bibfield  {author} {\bibinfo {author} {\bibfnamefont {M.}~\bibnamefont
  {{Onoe}}}\ and\ \bibinfo {author} {\bibfnamefont {M.}~\bibnamefont
  {{Kaneko}}},\ }\bibfield  {title} {\enquote {\bibinfo {title} {Computer
  generated pure binary hologram},}\ }\href@noop {} {\bibfield  {journal}
  {\bibinfo  {journal} {Electron. Commun. Jpn.}\ }\textbf {\bibinfo {volume}
  {62}},\ \bibinfo {pages} {118} (\bibinfo {year} {1979})}\BibitemShut
  {NoStop}%
\bibitem [{\citenamefont {Fujita}\ \emph {et~al.}(1996)\citenamefont {Fujita},
  \citenamefont {Morinaga}, \citenamefont {Kishimoto}, \citenamefont {Yasuda},
  \citenamefont {Matsui},\ and\ \citenamefont {Shimizu}}]{Fujita96}%
  \BibitemOpen
  \bibfield  {author} {\bibinfo {author} {\bibfnamefont {J.}~\bibnamefont
  {Fujita}}, \bibinfo {author} {\bibfnamefont {M.}~\bibnamefont {Morinaga}},
  \bibinfo {author} {\bibfnamefont {T.}~\bibnamefont {Kishimoto}}, \bibinfo
  {author} {\bibfnamefont {M.}~\bibnamefont {Yasuda}}, \bibinfo {author}
  {\bibfnamefont {S.}~\bibnamefont {Matsui}}, \ and\ \bibinfo {author}
  {\bibfnamefont {F.}~\bibnamefont {Shimizu}},\ }\bibfield  {title} {\enquote
  {\bibinfo {title} {Manipulation of an atomic beam by a computer-generated
  hologram},}\ }\href {\doibase 10.1038/380691a0} {\bibfield  {journal}
  {\bibinfo  {journal} {Nature}\ }\textbf {\bibinfo {volume} {380}},\ \bibinfo
  {pages} {691} (\bibinfo {year} {1996})}\BibitemShut {NoStop}%
\bibitem [{\citenamefont {Manfrinato}\ \emph {et~al.}(2013)\citenamefont
  {Manfrinato}, \citenamefont {Zhang}, \citenamefont {Su}, \citenamefont
  {Duan}, \citenamefont {Hobbs}, \citenamefont {Stach},\ and\ \citenamefont
  {Berggren}}]{Manfrinato13}%
  \BibitemOpen
  \bibfield  {author} {\bibinfo {author} {\bibfnamefont {V.~R.}\ \bibnamefont
  {Manfrinato}}, \bibinfo {author} {\bibfnamefont {L.}~\bibnamefont {Zhang}},
  \bibinfo {author} {\bibfnamefont {D.}~\bibnamefont {Su}}, \bibinfo {author}
  {\bibfnamefont {H.}~\bibnamefont {Duan}}, \bibinfo {author} {\bibfnamefont
  {R.~G.}\ \bibnamefont {Hobbs}}, \bibinfo {author} {\bibfnamefont {E.~A.}\
  \bibnamefont {Stach}}, \ and\ \bibinfo {author} {\bibfnamefont {K.~K.}\
  \bibnamefont {Berggren}},\ }\bibfield  {title} {\enquote {\bibinfo {title}
  {{Resolution Limits of Electron-Beam Lithography toward the Atomic Scale}},}\
  }\href {\doibase 10.1021/nl304715p} {\bibfield  {journal} {\bibinfo
  {journal} {Nano Lett.}\ }\textbf {\bibinfo {volume} {13}},\ \bibinfo {pages}
  {1555} (\bibinfo {year} {2013})}\BibitemShut {NoStop}%
\bibitem [{\citenamefont {Eder}\ \emph {et~al.}(2012)\citenamefont {Eder},
  \citenamefont {Reisinger}, \citenamefont {Greve}, \citenamefont {Bracco},\
  and\ \citenamefont {Holst}}]{Eder12}%
  \BibitemOpen
  \bibfield  {author} {\bibinfo {author} {\bibfnamefont {S.~D.}\ \bibnamefont
  {Eder}}, \bibinfo {author} {\bibfnamefont {T.}~\bibnamefont {Reisinger}},
  \bibinfo {author} {\bibfnamefont {M.~M.}\ \bibnamefont {Greve}}, \bibinfo
  {author} {\bibfnamefont {G.}~\bibnamefont {Bracco}}, \ and\ \bibinfo {author}
  {\bibfnamefont {B.}~\bibnamefont {Holst}},\ }\bibfield  {title} {\enquote
  {\bibinfo {title} {Focusing of a neutral helium beam below one micron},}\
  }\href {http://stacks.iop.org/1367-2630/14/i=7/a=073014} {\bibfield
  {journal} {\bibinfo  {journal} {New J. Phys.}\ }\textbf {\bibinfo {volume}
  {14}},\ \bibinfo {pages} {073014} (\bibinfo {year} {2012})}\BibitemShut
  {NoStop}%
\bibitem [{\citenamefont {Eder}\ \emph {et~al.}(2015)\citenamefont {Eder},
  \citenamefont {Guo}, \citenamefont {Kaltenbacher}, \citenamefont {Greve},
  \citenamefont {Kall\"ane}, \citenamefont {Kipp},\ and\ \citenamefont
  {Holst}}]{Eder15}%
  \BibitemOpen
  \bibfield  {author} {\bibinfo {author} {\bibfnamefont {S.~D.}\ \bibnamefont
  {Eder}}, \bibinfo {author} {\bibfnamefont {X.}~\bibnamefont {Guo}}, \bibinfo
  {author} {\bibfnamefont {T.}~\bibnamefont {Kaltenbacher}}, \bibinfo {author}
  {\bibfnamefont {M.~M.}\ \bibnamefont {Greve}}, \bibinfo {author}
  {\bibfnamefont {M.}~\bibnamefont {Kall\"ane}}, \bibinfo {author}
  {\bibfnamefont {L.}~\bibnamefont {Kipp}}, \ and\ \bibinfo {author}
  {\bibfnamefont {B.}~\bibnamefont {Holst}},\ }\bibfield  {title} {\enquote
  {\bibinfo {title} {Focusing of a neutral helium beam with a photon-sieve
  structure},}\ }\href {\doibase 10.1103/PhysRevA.91.043608} {\bibfield
  {journal} {\bibinfo  {journal} {Phys. Rev. A}\ }\textbf {\bibinfo {volume}
  {91}},\ \bibinfo {pages} {043608} (\bibinfo {year} {2015})}\BibitemShut
  {NoStop}%
\bibitem [{\citenamefont {Brand}\ \emph {et~al.}(2015)\citenamefont {Brand},
  \citenamefont {Sclafani}, \citenamefont {Knobloch}, \citenamefont {Lilach},
  \citenamefont {Juffmann}, \citenamefont {Kotakoski}, \citenamefont {Mangler},
  \citenamefont {Winter}, \citenamefont {Turchanin}, \citenamefont {Meyer},
  \citenamefont {Cheshnovsky},\ and\ \citenamefont {Arndt}}]{Arndt15}%
  \BibitemOpen
  \bibfield  {author} {\bibinfo {author} {\bibfnamefont {C.}~\bibnamefont
  {Brand}}, \bibinfo {author} {\bibfnamefont {M.}~\bibnamefont {Sclafani}},
  \bibinfo {author} {\bibfnamefont {C.}~\bibnamefont {Knobloch}}, \bibinfo
  {author} {\bibfnamefont {Y.}~\bibnamefont {Lilach}}, \bibinfo {author}
  {\bibfnamefont {T.}~\bibnamefont {Juffmann}}, \bibinfo {author}
  {\bibfnamefont {J.}~\bibnamefont {Kotakoski}}, \bibinfo {author}
  {\bibfnamefont {C.}~\bibnamefont {Mangler}}, \bibinfo {author} {\bibfnamefont
  {A.}~\bibnamefont {Winter}}, \bibinfo {author} {\bibfnamefont
  {A.}~\bibnamefont {Turchanin}}, \bibinfo {author} {\bibfnamefont
  {J.}~\bibnamefont {Meyer}}, \bibinfo {author} {\bibfnamefont
  {O.}~\bibnamefont {Cheshnovsky}}, \ and\ \bibinfo {author} {\bibfnamefont
  {M.}~\bibnamefont {Arndt}},\ }\bibfield  {title} {\enquote {\bibinfo {title}
  {An atomically thin matter-wave beamsplitter},}\ }\href {\doibase
  10.1038/nnano.2015.179} {\bibfield  {journal} {\bibinfo  {journal} {Nat.
  Nano}\ }\textbf {\bibinfo {volume} {10}},\ \bibinfo {pages} {845} (\bibinfo
  {year} {2015})}\BibitemShut {NoStop}%
\bibitem [{\citenamefont {Cronin}\ \emph {et~al.}(2009)\citenamefont {Cronin},
  \citenamefont {Schmiedmayer},\ and\ \citenamefont {Pritchard}}]{Pritchard09}%
  \BibitemOpen
  \bibfield  {author} {\bibinfo {author} {\bibfnamefont {A.~D.}\ \bibnamefont
  {Cronin}}, \bibinfo {author} {\bibfnamefont {J.}~\bibnamefont
  {Schmiedmayer}}, \ and\ \bibinfo {author} {\bibfnamefont {D.~E.}\
  \bibnamefont {Pritchard}},\ }\bibfield  {title} {\enquote {\bibinfo {title}
  {Optics and interferometry with atoms and molecules},}\ }\href {\doibase
  10.1103/RevModPhys.81.1051} {\bibfield  {journal} {\bibinfo  {journal} {Rev.
  Mod. Phys.}\ }\textbf {\bibinfo {volume} {81}},\ \bibinfo {pages} {1051}
  (\bibinfo {year} {2009})}\BibitemShut {NoStop}%
\bibitem [{\citenamefont {Nesse}\ \emph {et~al.}(2017)\citenamefont {Nesse},
  \citenamefont {Eder}, \citenamefont {Kaltenbacher}, \citenamefont {Grepstad},
  \citenamefont {Simonsen},\ and\ \citenamefont {Holst}}]{Nesse17}%
  \BibitemOpen
  \bibfield  {author} {\bibinfo {author} {\bibfnamefont {T.}~\bibnamefont
  {Nesse}}, \bibinfo {author} {\bibfnamefont {S.~D.}\ \bibnamefont {Eder}},
  \bibinfo {author} {\bibfnamefont {T.}~\bibnamefont {Kaltenbacher}}, \bibinfo
  {author} {\bibfnamefont {J.~O.}\ \bibnamefont {Grepstad}}, \bibinfo {author}
  {\bibfnamefont {I.}~\bibnamefont {Simonsen}}, \ and\ \bibinfo {author}
  {\bibfnamefont {B.}~\bibnamefont {Holst}},\ }\bibfield  {title} {\enquote
  {\bibinfo {title} {Neutral-helium-atom diffraction from a micron-scale
  periodic structure: Photonic-crystal-membrane characterization},}\ }\href
  {\doibase 10.1103/PhysRevA.95.063618} {\bibfield  {journal} {\bibinfo
  {journal} {Phys. Rev. A}\ }\textbf {\bibinfo {volume} {95}},\ \bibinfo
  {pages} {063618} (\bibinfo {year} {2017})}\BibitemShut {NoStop}%
\bibitem [{\citenamefont {Patton}\ \emph {et~al.}(2006)\citenamefont {Patton},
  \citenamefont {Deponte}, \citenamefont {Elliott},\ and\ \citenamefont
  {Kevan}}]{Patton2006}%
  \BibitemOpen
  \bibfield  {author} {\bibinfo {author} {\bibfnamefont {F.~S.}\ \bibnamefont
  {Patton}}, \bibinfo {author} {\bibfnamefont {D.~P.}\ \bibnamefont {Deponte}},
  \bibinfo {author} {\bibfnamefont {G.~S.}\ \bibnamefont {Elliott}}, \ and\
  \bibinfo {author} {\bibfnamefont {S.~D.}\ \bibnamefont {Kevan}},\ }\bibfield
  {title} {\enquote {\bibinfo {title} {{Speckle Patterns with Atomic and
  Molecular de {Broglie} Waves}},}\ }\href {\doibase
  10.1103/PhysRevLett.97.013202} {\bibfield  {journal} {\bibinfo  {journal}
  {Phys. Rev. Lett.}\ }\textbf {\bibinfo {volume} {97}},\ \bibinfo {pages}
  {013202} (\bibinfo {year} {2006})}\BibitemShut {NoStop}%
\bibitem [{\citenamefont {Even}(2015)}]{Even2015}%
  \BibitemOpen
  \bibfield  {author} {\bibinfo {author} {\bibfnamefont {U.}~\bibnamefont
  {Even}},\ }\bibfield  {title} {\enquote {\bibinfo {title} {{The {Even-Lavie}
  valve as a source for high intensity supersonic beam}},}\ }\href {\doibase
  10.1140/epjti/s40485-015-0027-5} {\bibfield  {journal} {\bibinfo  {journal}
  {EPJ Tech. Instrum.}\ }\textbf {\bibinfo {volume} {2}},\ \bibinfo {pages}
  {17} (\bibinfo {year} {2015})}\BibitemShut {NoStop}%
\bibitem [{\citenamefont {Robert}\ \emph {et~al.}(2001)\citenamefont {Robert},
  \citenamefont {Sirjean}, \citenamefont {Browaeys}, \citenamefont {Poupard},
  \citenamefont {Nowak}, \citenamefont {Boiron}, \citenamefont {Westbrook},\
  and\ \citenamefont {Aspect}}]{Robert01}%
  \BibitemOpen
  \bibfield  {author} {\bibinfo {author} {\bibfnamefont {A.}~\bibnamefont
  {Robert}}, \bibinfo {author} {\bibfnamefont {O.}~\bibnamefont {Sirjean}},
  \bibinfo {author} {\bibfnamefont {A.}~\bibnamefont {Browaeys}}, \bibinfo
  {author} {\bibfnamefont {J.}~\bibnamefont {Poupard}}, \bibinfo {author}
  {\bibfnamefont {S.}~\bibnamefont {Nowak}}, \bibinfo {author} {\bibfnamefont
  {D.}~\bibnamefont {Boiron}}, \bibinfo {author} {\bibfnamefont {C.~I.}\
  \bibnamefont {Westbrook}}, \ and\ \bibinfo {author} {\bibfnamefont
  {A.}~\bibnamefont {Aspect}},\ }\bibfield  {title} {\enquote {\bibinfo {title}
  {{A {Bose-Einstein} Condensate of Metastable Atoms}},}\ }\href {\doibase
  10.1126/science.1060622} {\bibfield  {journal} {\bibinfo  {journal}
  {Science}\ }\textbf {\bibinfo {volume} {292}},\ \bibinfo {pages} {461}
  (\bibinfo {year} {2001})}\BibitemShut {NoStop}%
\bibitem [{\citenamefont {Pereira Dos~Santos}\ \emph
  {et~al.}(2001)\citenamefont {Pereira Dos~Santos}, \citenamefont {L\'eonard},
  \citenamefont {Wang}, \citenamefont {Barrelet}, \citenamefont {Perales},
  \citenamefont {Rasel}, \citenamefont {Unnikrishnan}, \citenamefont {Leduc},\
  and\ \citenamefont {Cohen-Tannoudji}}]{Santos01}%
  \BibitemOpen
  \bibfield  {author} {\bibinfo {author} {\bibfnamefont {F.}~\bibnamefont
  {Pereira Dos~Santos}}, \bibinfo {author} {\bibfnamefont {J.}~\bibnamefont
  {L\'eonard}}, \bibinfo {author} {\bibfnamefont {J.}~\bibnamefont {Wang}},
  \bibinfo {author} {\bibfnamefont {C.~J.}\ \bibnamefont {Barrelet}}, \bibinfo
  {author} {\bibfnamefont {F.}~\bibnamefont {Perales}}, \bibinfo {author}
  {\bibfnamefont {E.}~\bibnamefont {Rasel}}, \bibinfo {author} {\bibfnamefont
  {C.~S.}\ \bibnamefont {Unnikrishnan}}, \bibinfo {author} {\bibfnamefont
  {M.}~\bibnamefont {Leduc}}, \ and\ \bibinfo {author} {\bibfnamefont
  {C.}~\bibnamefont {Cohen-Tannoudji}},\ }\bibfield  {title} {\enquote
  {\bibinfo {title} {{{Bose-Einstein} Condensation of Metastable Helium}},}\
  }\href {\doibase 10.1103/PhysRevLett.86.3459} {\bibfield  {journal} {\bibinfo
   {journal} {Phys. Rev. Lett.}\ }\textbf {\bibinfo {volume} {86}},\ \bibinfo
  {pages} {3459} (\bibinfo {year} {2001})}\BibitemShut {NoStop}%
\bibitem [{\citenamefont {Keller}\ \emph {et~al.}(2014)\citenamefont {Keller},
  \citenamefont {Kotyrba}, \citenamefont {Leupold}, \citenamefont {Singh},
  \citenamefont {Ebner},\ and\ \citenamefont {Zeilinger}}]{Zeilinger14}%
  \BibitemOpen
  \bibfield  {author} {\bibinfo {author} {\bibfnamefont {M.}~\bibnamefont
  {Keller}}, \bibinfo {author} {\bibfnamefont {M.}~\bibnamefont {Kotyrba}},
  \bibinfo {author} {\bibfnamefont {F.}~\bibnamefont {Leupold}}, \bibinfo
  {author} {\bibfnamefont {M.}~\bibnamefont {Singh}}, \bibinfo {author}
  {\bibfnamefont {M.}~\bibnamefont {Ebner}}, \ and\ \bibinfo {author}
  {\bibfnamefont {A.}~\bibnamefont {Zeilinger}},\ }\bibfield  {title} {\enquote
  {\bibinfo {title} {{{Bose-Einstein} condensate of metastable helium for
  quantum correlation experiments}},}\ }\href {\doibase
  10.1103/PhysRevA.90.063607} {\bibfield  {journal} {\bibinfo  {journal} {Phys.
  Rev. A}\ }\textbf {\bibinfo {volume} {90}},\ \bibinfo {pages} {063607}
  (\bibinfo {year} {2014})}\BibitemShut {NoStop}%
\bibitem [{\citenamefont {Fouda}\ \emph {et~al.}(2016)\citenamefont {Fouda},
  \citenamefont {Fang}, \citenamefont {Ketterson},\ and\ \citenamefont
  {Shahriar}}]{Fouda2016}%
  \BibitemOpen
  \bibfield  {author} {\bibinfo {author} {\bibfnamefont {M.~F.}\ \bibnamefont
  {Fouda}}, \bibinfo {author} {\bibfnamefont {R.}~\bibnamefont {Fang}},
  \bibinfo {author} {\bibfnamefont {J.~B.}\ \bibnamefont {Ketterson}}, \ and\
  \bibinfo {author} {\bibfnamefont {M.~S.}\ \bibnamefont {Shahriar}},\
  }\bibfield  {title} {\enquote {\bibinfo {title} {Generation of arbitrary
  lithographic patterns using {Bose-Einstein-condensate} interferometry},}\
  }\href {\doibase 10.1103/PhysRevA.94.063644} {\bibfield  {journal} {\bibinfo
  {journal} {Phys. Rev. A}\ }\textbf {\bibinfo {volume} {94}},\ \bibinfo
  {pages} {063644} (\bibinfo {year} {2016})}\BibitemShut {NoStop}%
\bibitem [{\citenamefont {Eder}\ \emph {et~al.}(2017)\citenamefont {Eder},
  \citenamefont {Ravn}, \citenamefont {Samelin}, \citenamefont {Bracco},
  \citenamefont {Palau}, \citenamefont {Reisinger}, \citenamefont {Knudsen},
  \citenamefont {Lefmann},\ and\ \citenamefont {Holst}}]{Eder2017}%
  \BibitemOpen
  \bibfield  {author} {\bibinfo {author} {\bibfnamefont {S.~D.}\ \bibnamefont
  {Eder}}, \bibinfo {author} {\bibfnamefont {A.~K.}\ \bibnamefont {Ravn}},
  \bibinfo {author} {\bibfnamefont {B.}~\bibnamefont {Samelin}}, \bibinfo
  {author} {\bibfnamefont {G.}~\bibnamefont {Bracco}}, \bibinfo {author}
  {\bibfnamefont {A.~S.}\ \bibnamefont {Palau}}, \bibinfo {author}
  {\bibfnamefont {T.}~\bibnamefont {Reisinger}}, \bibinfo {author}
  {\bibfnamefont {E.~B.}\ \bibnamefont {Knudsen}}, \bibinfo {author}
  {\bibfnamefont {K.}~\bibnamefont {Lefmann}}, \ and\ \bibinfo {author}
  {\bibfnamefont {B.}~\bibnamefont {Holst}},\ }\bibfield  {title} {\enquote
  {\bibinfo {title} {Zero-order filter for diffractive focusing of de {Broglie}
  matter waves},}\ }\href {\doibase 10.1103/PhysRevA.95.023618} {\bibfield
  {journal} {\bibinfo  {journal} {Phys. Rev. A}\ }\textbf {\bibinfo {volume}
  {95}},\ \bibinfo {pages} {023618} (\bibinfo {year} {2017})}\BibitemShut
  {NoStop}%
\bibitem [{\citenamefont {Koch}\ \emph {et~al.}(2008)\citenamefont {Koch},
  \citenamefont {Rehbein}, \citenamefont {Schmahl}, \citenamefont {Reisinger},
  \citenamefont {Bracco}, \citenamefont {Ernst},\ and\ \citenamefont
  {Holst}}]{Koch2008}%
  \BibitemOpen
  \bibfield  {author} {\bibinfo {author} {\bibfnamefont {M.}~\bibnamefont
  {Koch}}, \bibinfo {author} {\bibfnamefont {S.}~\bibnamefont {Rehbein}},
  \bibinfo {author} {\bibfnamefont {G.}~\bibnamefont {Schmahl}}, \bibinfo
  {author} {\bibfnamefont {T.}~\bibnamefont {Reisinger}}, \bibinfo {author}
  {\bibfnamefont {G.}~\bibnamefont {Bracco}}, \bibinfo {author} {\bibfnamefont
  {W.~E.}\ \bibnamefont {Ernst}}, \ and\ \bibinfo {author} {\bibfnamefont
  {B.}~\bibnamefont {Holst}},\ }\bibfield  {title} {\enquote {\bibinfo {title}
  {{Imaging with neutral atoms -- a new matter-wave microscope}},}\ }\href
  {\doibase 10.1111/j.1365-2818.2007.01874.x} {\bibfield  {journal} {\bibinfo
  {journal} {J. Microsc.}\ }\textbf {\bibinfo {volume} {229}},\ \bibinfo
  {pages} {1} (\bibinfo {year} {2008})}\BibitemShut {NoStop}%
\bibitem [{\citenamefont {Doak}\ \emph {et~al.}(1999)\citenamefont {Doak},
  \citenamefont {Grisenti}, \citenamefont {Rehbein}, \citenamefont {Schmahl},
  \citenamefont {Toennies},\ and\ \citenamefont {W\"oll}}]{Doak1999}%
  \BibitemOpen
  \bibfield  {author} {\bibinfo {author} {\bibfnamefont {R.~B.}\ \bibnamefont
  {Doak}}, \bibinfo {author} {\bibfnamefont {R.~E.}\ \bibnamefont {Grisenti}},
  \bibinfo {author} {\bibfnamefont {S.}~\bibnamefont {Rehbein}}, \bibinfo
  {author} {\bibfnamefont {G.}~\bibnamefont {Schmahl}}, \bibinfo {author}
  {\bibfnamefont {J.~P.}\ \bibnamefont {Toennies}}, \ and\ \bibinfo {author}
  {\bibfnamefont {C.}~\bibnamefont {W\"oll}},\ }\bibfield  {title} {\enquote
  {\bibinfo {title} {{Towards Realization of an Atomic de {Broglie} Microscope:
  Helium Atom Focusing Using {Fresnel} Zone Plates}},}\ }\href {\doibase
  10.1103/PhysRevLett.83.4229} {\bibfield  {journal} {\bibinfo  {journal}
  {Phys. Rev. Lett.}\ }\textbf {\bibinfo {volume} {83}},\ \bibinfo {pages}
  {4229} (\bibinfo {year} {1999})}\BibitemShut {NoStop}%
\bibitem [{\citenamefont {Carnal}\ \emph {et~al.}(1991)\citenamefont {Carnal},
  \citenamefont {Sigel}, \citenamefont {Sleator}, \citenamefont {Takuma},\ and\
  \citenamefont {Mlynek}}]{Carnal1991}%
  \BibitemOpen
  \bibfield  {author} {\bibinfo {author} {\bibfnamefont {O.}~\bibnamefont
  {Carnal}}, \bibinfo {author} {\bibfnamefont {M.}~\bibnamefont {Sigel}},
  \bibinfo {author} {\bibfnamefont {T.}~\bibnamefont {Sleator}}, \bibinfo
  {author} {\bibfnamefont {H.}~\bibnamefont {Takuma}}, \ and\ \bibinfo {author}
  {\bibfnamefont {J.}~\bibnamefont {Mlynek}},\ }\bibfield  {title} {\enquote
  {\bibinfo {title} {Imaging and focusing of atoms by a fresnel zone plate},}\
  }\href {\doibase 10.1103/PhysRevLett.67.3231} {\bibfield  {journal} {\bibinfo
   {journal} {Phys. Rev. Lett.}\ }\textbf {\bibinfo {volume} {67}},\ \bibinfo
  {pages} {3231} (\bibinfo {year} {1991})}\BibitemShut {NoStop}%
\bibitem [{\citenamefont {Gardner}\ \emph {et~al.}(2017)\citenamefont
  {Gardner}, \citenamefont {Anciaux},\ and\ \citenamefont
  {Raizen}}]{Gardner2017}%
  \BibitemOpen
  \bibfield  {author} {\bibinfo {author} {\bibfnamefont {J.~R.}\ \bibnamefont
  {Gardner}}, \bibinfo {author} {\bibfnamefont {E.~M.}\ \bibnamefont
  {Anciaux}}, \ and\ \bibinfo {author} {\bibfnamefont {M.~G.}\ \bibnamefont
  {Raizen}},\ }\bibfield  {title} {\enquote {\bibinfo {title} {Communication:
  Neutral atom imaging using a pulsed electromagnetic lens},}\ }\href {\doibase
  10.1063/1.4976986} {\bibfield  {journal} {\bibinfo  {journal} {J. Chem.
  Phys.}\ }\textbf {\bibinfo {volume} {146}},\ \bibinfo {pages} {081102}
  (\bibinfo {year} {2017})}\BibitemShut {NoStop}%
\bibitem [{\citenamefont {Goodman}(2005)}]{Goodman05}%
  \BibitemOpen
  \bibfield  {author} {\bibinfo {author} {\bibfnamefont {J.}~\bibnamefont
  {Goodman}},\ }\href {https://books.google.no/books?id=ow5xs\_Rtt9AC} {\emph
  {\bibinfo {title} {{Introduction to {Fourier} Optics}}}},\ McGraw-Hill
  physical and quantum electronics series\ (\bibinfo  {publisher} {W. H.
  Freeman},\ \bibinfo {year} {2005})\BibitemShut {NoStop}%
\bibitem [{\citenamefont {Adams}\ \emph {et~al.}(1994)\citenamefont {Adams},
  \citenamefont {Sigel},\ and\ \citenamefont {Mlynek}}]{Adams94}%
  \BibitemOpen
  \bibfield  {author} {\bibinfo {author} {\bibfnamefont {C.~S.}\ \bibnamefont
  {Adams}}, \bibinfo {author} {\bibfnamefont {M.}~\bibnamefont {Sigel}}, \ and\
  \bibinfo {author} {\bibfnamefont {J.}~\bibnamefont {Mlynek}},\ }\bibfield
  {title} {\enquote {\bibinfo {title} {Atom optics},}\ }\href {\doibase
  http://dx.doi.org/10.1016/0370-1573(94)90066-3} {\bibfield  {journal}
  {\bibinfo  {journal} {Phys. Rep.}\ }\textbf {\bibinfo {volume} {240}},\
  \bibinfo {pages} {143 } (\bibinfo {year} {1994})}\BibitemShut {NoStop}%
\bibitem [{\citenamefont {Press}\ \emph {et~al.}(2007)\citenamefont {Press},
  \citenamefont {Teukolsky}, \citenamefont {Vetterling},\ and\ \citenamefont
  {Flannery}}]{Book:NumericalRecipies}%
  \BibitemOpen
  \bibfield  {author} {\bibinfo {author} {\bibfnamefont {W.~H.}\ \bibnamefont
  {Press}}, \bibinfo {author} {\bibfnamefont {S.~A.}\ \bibnamefont
  {Teukolsky}}, \bibinfo {author} {\bibfnamefont {W.~T.}\ \bibnamefont
  {Vetterling}}, \ and\ \bibinfo {author} {\bibfnamefont {B.~P.}\ \bibnamefont
  {Flannery}},\ }\href@noop {} {\emph {\bibinfo {title} {Numerical Recipes 3rd
  Edition: The Art of Scientific Computing}}},\ \bibinfo {edition} {3rd}\ ed.\
  (\bibinfo  {publisher} {Cambridge University Press},\ \bibinfo {address} {New
  York, NY, USA},\ \bibinfo {year} {2007})\BibitemShut {NoStop}%
\bibitem [{Note1()}]{Note1}%
  \BibitemOpen
  \bibinfo {note} {It is easy to extended the procedure to the case of a
  different subdivision along the two axes, but this will not be discussed
  here.}\BibitemShut {Stop}%
\end{thebibliography}%

\end{document}